\documentclass[a4paper]{paper_cw}

\usepackage{helvet}
\usepackage{amsmath}
\usepackage{amssymb}
\usepackage{mathrsfs}
\usepackage{graphics}
\usepackage[labelformat=simple]{subfig}
\usepackage{marvosym}
\usepackage[numbers]{natbib}
\usepackage{hyperref}
\usepackage{siunitx}
\usepackage{algorithm}
\usepackage{algpseudocode}
\usepackage{booktabs}
\usepackage{xcolor}
\usepackage{float}
\usepackage{graphicx}
\usepackage[page]{appendix}

\usepackage{lineno}
\makeatletter
\renewcommand{\fnum@figure}{\textbf{Fig. \thefigure}}
\makeatother

 \newcommand{\Rev}{\textcolor{black}}

\long\def\symbolfootnote[#1]#2{\begingroup \def\thefootnote{\fnsymbol{footnote}}\footnote[#1]{#2} \endgroup}

\renewcommand{\vec}[1]{ \ensuremath{ \mathbf{ #1 } } }

\newcommand{\gvec}[1]{ \ensuremath{ \boldsymbol{ #1 } } }

\newcommand{\de}{\,\mbox{det}\,}

\newcommand{\calK}{\ensuremath{ \mathcal{K} }}

\newcommand{\calO}{\ensuremath{ \mathcal{O} }}

\newcommand{\bbN}{\ensuremath{ \mathbb{N} }}

\newcommand{\bbP}{\ensuremath{ \mathbb{P} }}

\newcommand{\bbR}{\ensuremath{ \mathbb{R} }}

\newcommand{\vecb}{\ensuremath{ \vec{b} }}
\newcommand{\vecc}{\ensuremath{ \vec{c} }}
\newcommand{\vecd}{\ensuremath{ \vec{d} }}
\newcommand{\vece}{\ensuremath{ \vec{e} }}

\newcommand{\vecu}{\ensuremath{ \vec{u} }}

\newcommand{\vecw}{\ensuremath{ \vec{w} }}

\newcommand{\vecy}{\ensuremath{ \vec{y} }}
\newcommand{\vecz}{\ensuremath{ \vec{z} }}

\newcommand{\vecA}{\ensuremath{ \vec{A} }}

\newcommand{\vecC}{\ensuremath{ \vec{C} }}
\newcommand{\vecD}{\ensuremath{ \vec{D} }}
\newcommand{\vecE}{\ensuremath{ \vec{E} }}
\newcommand{\vecF}{\ensuremath{ \vec{F} }}

\newcommand{\vecH}{\ensuremath{ \vec{H} }}
\newcommand{\vecI}{\ensuremath{ \vec{I} }}

\newcommand{\vecK}{\ensuremath{ \vec{K} }}

\newcommand{\vecN}{\ensuremath{ \vec{N} }}

\newcommand{\vecP}{\ensuremath{ \vec{P} }}

\newcommand{\vecS}{\ensuremath{ \vec{S} }}
\newcommand{\vecT}{\ensuremath{ \vec{T} }}

\newcommand{\vecX}{\ensuremath{ \vec{X} }}
\newcommand{\vecY}{\ensuremath{ \vec{Y} }}

\newcommand{\vecepsilon   }{\ensuremath{ \gvec{\epsilon} }}

\newcommand{\vecvartheta  }{\ensuremath{ \gvec{\vartheta} }}

\newcommand{\vecmu        }{\ensuremath{ \gvec{\mu} }}

\newcommand{\vecxi        }{\ensuremath{ \gvec{\xi} }}

\newcommand{\vecsigma     }{\ensuremath{ \gvec{\sigma} }}

\newcommand{\onetwo}{\frac{1}{2}}

\renewcommand{\d}[1]{\text{$\hspace{0.1cm}$d $\hspace{-0.11cm}#1$}}
\newcommand{\del}{\ensuremath{\partial}}

\newcommand{\grad}[1]{\ensuremath{ \boldsymbol{\nabla}{#1}}}

\newcommand{\parder}[2]{\ensuremath{ \frac{\del #1}{\del #2} }}

\renewcommand{\ln}[1]{\text{$\hspace{0.1cm}$ln$\left(#1\right)$}}
\renewcommand{\exp}[1]{\ensuremath{ \,\text{exp}{\left( #1 \right)} }}
\newcommand{\expbig}[1]{\ensuremath{ \,\text{exp}{\big( #1 \big)} }}

\newcommand{\tr}[1]{\text{$\hspace{0.1cm}$tr$\left(#1\right)$}}

\newcommand{\percent}{\ensuremath{ \%  }}

\newcommand{\saSp}{\boldsymbol{\Theta}}                       
\newcommand{\sigAl}{\mathscr{F}}                              
\newcommand{\prb}{\bbP}                                       
\newcommand{\isGP}{\sim \mathcal{GP}}                         
\newcommand{\vecEmpty}{\boldsymbol{\varnothing}}              

\newcommand{\mPsi}{\boldsymbol{\Psi}}                         
\DeclareMathOperator*{\argmax}{arg\,max}                      
\DeclareMathOperator*{\argmin}{arg\,min}                      
\newcommand{\NMC}{N^{\text{MC}}}                              
\newcommand{\MuMC}{\vecmu_{\vecu^{\text{MC}}}}                
\newcommand{\MuPC}{\vecmu_{\vecu^{\text{PC}}}}                
\newcommand{\CuMC}{\vecC_{\vecu^{\text{MC}}}}                 
\newcommand{\CuPC}{\vecC_{\vecu^{\text{PC}}}}                 
\newcommand{\YObs}{\vecY}                        
\newcommand{\YObsi}{\vecy_i}                     
\newcommand{\yObs}{\vecy}                        
\newcommand{\yObsOne}{\vecy_1}                     
\newcommand{\yObsTwo}{\vecy_2}                     
\newcommand{\yObsNrep}{\vecy_{\nrep}}                     
\newcommand{\zRel}{\vecz}                       
\newcommand{\xsen}{\vecX_{\text{sen}}}                        
\newcommand{\nnode}{n_{\text{nod}}}                           
\newcommand{\nsen}{n_{\text{sen}}}                            
\newcommand{\ngdof}{n_{\text{gdof}}}                            
\newcommand{\ngsen}{n_{\text{gsen}}}                            
\newcommand{\nrep}{n_{\text{rep}}}                            
\newcommand{\CD}{\vecC_\vecd}                                 
\newcommand{\CE}{\vecC_\vece}                                 
\newcommand{\CZ}{\vecC_\vecz}                                 
\newcommand{\MuPCY}{\vecmu_{\vecu^{\text{PC}}|\YObs}}                     
\newcommand{\CuPCY}{\vecC_{\vecu^{\text{PC}}|\YObs}}                      
\newcommand{\Muz}{\vecmu_{\vecz}}                     
\newcommand{\sigd}{\sigma_{\vecd}}                            
\newcommand{\lsigd}{\ln{\sigd}}                               
\newcommand{\lsigdStr}{\ln{\sigd^*}}                               
\newcommand{\ld}{l_{\vecd}}                                   
\newcommand{\lld}{\ln{\ld}}                                   
\newcommand{\lldStr}{\ln{\ld^*}}                                   
\newcommand{\dcsigd}{c_{\vecd,\sigd}}                         
\newcommand{\dcld}{c_{\vecd,\ld}}                             
\newcommand{\dCsigd}{\vecC_{\vecd,\sigd}}                     
\newcommand{\dCld}{\vecC_{\vecd,\ld}}                         
\newcommand{\vecvarSigma}{\ensuremath{\gvec{\varSigma}}}

\newcommand{\MuxPC}{\vecmu_{\vecu_X^{\text{PC}}}}

\newcommand{\MsigmaxxPC}{\vecmu_{\vecsigma_{XX}^{\text{MC}}}}
\newcommand{\MeanPiolaPC}{\vecmu_{\vecP_{XX}^{\text{MC}}}}

\newcommand{\STDuxPC}{\textbf{\text{std}}_{\vecu_X^{\text{PC}}}}
\newcommand{\STDuxMC}{\textbf{\text{std}}_{\vecu_X^{\text{MC}}}}

\newcommand{\MuyPCY}{\vecmu_{\vecu_Y^{\text{PC}}|\YObs}}
\newcommand{\MuxPCY}{\vecmu_{\vecu_X^{\text{PC}}|\YObs}}

\newcommand{\SuxPCY}{\textbf{\text{std}}_{\vecu_X^{\text{PC}}|\YObs}}
\newcommand{\MsigmaxxPCY}{{\vecsigma_{XX}|\YObs}}
\newcommand{\MsigmaPCY}{{\vecsigma|\YObs}}

\newcommand{\MeanPiolaPCY}{{\vecP_{XX}|\YObs}}
\newcommand{\MPiolaPCY}{{\vecP|\YObs}}
\newcommand{\ModelLE}{{\mathcal{M}_{\text{LE}}}}
\newcommand{\ModelSV}{{\mathcal{M}_{\text{SV}}}}
\newcommand{\BFSVLE}{{BF_{\ModelSV,\ModelLE}}}

\begin{document}
\title{Inferring Displacement Fields from Sparse Measurements Using the Statistical Finite Element Method}
\author{Vahab {B. Narouie}$^{1}$ (\Letter), Henning {Wessels}$^{1}$, Ulrich {R{\"o}mer}$^{2}$}
\institute{(1) Institute of Computational Modeling in Civil Engineering, Technische Universität Braunschweig, Pockelsstr. 3, 38106 Braunschweig, Germany \\
	\noindent(2) Institut für Dynamik und Schwingungen, Technische Universität Braunschweig, Schleinitzstr. 20, 38106 Braunschweig, Germany \\
	\email{v.narouie@tu-braunschweig.de} \\
}
\maketitle
\thispagestyle{empty}
\abstract
{
	Nowadays, strain and displacement can be measured using techniques such as electronic speckle pattern interferometry and digital image correlation. However, usually, only some part of the domain of interest is accessible for measurement devices. In order to assess also inaccessible areas of the structure, a fundamental task for structural health monitoring is the inference of full-field displacements from sparse measurements. Another important task is the inference of mechanical stress from displacement or strain data. In computational mechanics, the link between such data and stress is established via constitutive models.

	A well-established approach for inferring full displacement and stress fields from possibly sparse data is to calibrate the parameter of a given constitutive model using a Bayesian update. After calibration, a (stochastic) forward simulation is conducted with the identified model parameters to resolve physical fields in regions that were not accessible to the measurement device.  A shortcoming of model calibration is that the model is deemed to best represent reality, which is only sometimes the case, especially in the context of the aging of structures and materials. While this issue is often addressed with repeated model calibration, a different approach is followed in the recently proposed statistical Finite Element Method (statFEM). Instead of using Bayes' theorem to update model parameters, the displacement is chosen as the stochastic prior and updated to fit the measurement data more closely. For this purpose, the statFEM framework introduces a so-called model-reality mismatch, parametrized by only three hyperparameters. This makes the inference of full-field data computationally efficient in an online stage: If the stochastic prior can be computed offline, solving the underlying partial differential equation (PDE) online is unnecessary. Compared to solving a PDE, identifying only three hyperparameters and conditioning the state on the sensor data requires much fewer computational resources. Computational efficiency is an essential requirement for online applications. It is noted that conditioning a model-based prior field on sensor data represents a variant of physics-based regression.

	This paper presents two contributions to the existing statFEM approach: First, we use a non-intrusive polynomial chaos method to compute the prior, enabling the use of complex mechanical models in deterministic formulations.  Second, we examine the influence of prior material models (linear elastic and St.Venant Kirchhoff material with uncertain Young's modulus) on the updated solution. We present statFEM results for $1$D and $2$D examples, while an extension to $3$D is straightforward.
}
\keywords{Stochastic Finite Element Method \and Bayesian Calibration \and Polynomial Chaos Expansion \and Non-linear Elasticity \and Statistical Finite Element Method \and Physics-constrained Regression}

\section{Introduction}\label{sec:introduction}
Today, there is a significant interest in adopting emerging sensing technologies for instrumentation within various structural systems. For example, wireless sensors, digital video cameras, and sensor networks are emerging as sensing paradigms that collect data for their in situ monitoring \cite{xiao2020development}. Moreover, digital image correlation (DIC) is a well-known technique that measures the surface displacement between initial and deformed states \cite{sutton2009image,rastogi2013optical,sutton2013computer}. On the other hand, studying physical systems based on models has been essential to science, engineering, and industry for many years, mathematically described using PDEs. Efficient and accurate solution methods for PDEs are an impetus of the computational engineering research community. For example, the \textit{Finite Element Method} (FEM) \cite{strang2018analysis} and \textit{Isogeometric Analysis} (IGA) \cite{Hughes2005} schemes approximate the solution by solving a discretized problem on a finite mesh of the PDE domain. These computational methods are widely applied in fluid and structural mechanics, wave propagation, and heat conduction, among many other areas \cite{krysl2006pragmatic,cottrell2009isogeometric}.

In computational mechanics, one crucial aspect is choosing or even formulating a constitutive model with a priori unknown material parameters. Inferring material parameters of a constitutive model given observation data is called parameter identification, which can be considered an inverse problem \cite{tarantola2005inverse}. A widespread method for identifying the material parameters is the minimization of the least squares error between the observed experimental data and the model response, see, e.g., \cite{Mahnken2017}. In \cite{schnur1992inverse}, the usage of FEM to identify elastic materials with inclusion is demonstrated, and some further contributions are provided in \cite{benedix1998local,cooreman2007elasto}. Such an approach offers a deterministic estimate of material parameters. Alternatively, \textit{Physics Informed Neural Networks} (PINNs) can be used for parameter identification. Recently, a PINNs formulation to identify material parameters in a realistic data regime is represented in \cite{anton2022}. The linear elastic and elastoplastic parameter identification process with PINNs is also described in \cite{haghighat2021physics}, and other contributions are provided in \cite{zhang2020physics,hamel2022calibrating}.

However, the observation data are corrupted in almost any circumstance. This corruption may be caused by measurement noise and imprecise devices. Modeling assumptions are another issue. Model inadequacies are intrinsic in any physical model up to a certain level, e.g., due to missing physics or unresolved scales. Hence, both data and model can only represent reality up to a certain level of accuracy and uncertainty. Investigating different sources and levels of inaccuracy, variability, and numerical and modeling errors is the objective of \textit{uncertainty quantification} (UQ). Therefore, a significant shortcoming of the aforementioned least squares approach is that it lacks any statement about the associated uncertainty. An alternative approach is \textit{Bayesian inference} (BI). BI combines prior knowledge about the unknown material parameters with the knowledge learned from observation data to determine the posterior distribution of the unknown material parameters. In other words, BI conditions the assumed prior knowledge about the unknown material parameters on the observation data \cite{isenberg1979progressing,alvin1997finite,marwala2005finite,koutsourelakis2012novel}. In the context of biomechanics, a Bayesian approach for parameter estimation with quantified uncertainties is presented in \cite{romer2022surrogate} and for selecting the best hyperelastic material model for soft tissues in \cite{madireddy2015bayesiain}.

Furthermore, reconstructing the full displacement, strain, and stress fields from observation data is essential in many engineering applications. Fully resolved physical fields can be used to inform \Rev{\textit{structural health monitoring} (SHM)} and consequently assist, e.g., with the maintenance of structures. \Rev{One method for this purpose is the \textit{inverse finite element method} (iFEM) \cite{tessler2003variational,tessler2005least}. The iFEM involves minimizing a weighted least-squares functional that measures the discrepancy between the strains calculated from reconstructed displacements using finite elements and the measured strains. Minimizing the functional gives rise to a system of algebraic equations, where the nodal degrees of freedom of the finite element mesh serve as the unknowns. Upon obtaining the nodal unknowns, the displacement field can be fully reconstructed through the utilization of the corresponding shape functions. This approach relies solely on strain-displacement relations and requires no information about the materials or load applied to the structure. The iFEM has been developed and applied to various structures, including $1$D beams and frames and $2$D plates and shells \cite{tessler2004inverse, gherlone2012shape} and has been demonstrated to be effective for SHM on simple beams using fiber optic strain measurements \cite{quach2005structural}. The work presented in \cite{gherlone2018shape} demonstrates the application of iFEM for reconstructing the full-field displacements of a real wing-shaped thin aluminum plate. Another technique is the Modal Method (MM), which represents the displacements and strains in terms of modal shapes and modal coordinates. MM has been used to predict the static deformation of a cantilevered aluminum plate by utilizing experimentally measured modal characteristics in \cite{foss1995using}.
The conventional iFEM and MM neglect inherent uncertainty in material parameters, which lacks any statement about the associated uncertainty impacted on reconstructed displacement. However, a recent study by Esposito et al. \cite{esposito2021material} demonstrates that material uncertainties and noisy strain measurements can impact the reconstructed displacement field. The authors evaluated the robustness of iFEM and MM with respect to uncertain inputs using \textit{Latin Hypercube Sampling} (LHS). In addition, a stochastic variant of iFEM has been introduced recently in \cite{poloni2023towards}, utilizing a Gaussian Process as a pre-extrapolation and interpolation method for strains. This enables using confidence intervals as a metric to set iFEM weights through a mapping process, thereby enabling a more coherent approach than the deterministic method that relies on arbitrary weight assignment. Furthermore, the interpolation/extrapolation uncertainty permits the computation of the uncertainty of the iFEM displacements, resulting in a non-deterministic iFEM solution. } Alternatively, surrogate models can be developed that map observational data to full fields. While those are much more efficient to evaluate, surrogates such as neural networks \cite{erichson2020shallow} require a significant amount of training data and may have limited generalization capability.

In this contribution, we follow a \Rev{probabilistic} approach, recently proposed as the \textit{statistical finite element method}  (statFEM) \cite{girolami2021statistical}. This approach is also based on BI but with a significant difference. Instead of updating the model parameter, the PDE solution is updated directly. Hence, statFEM combines prior knowledge about the displacement field with the knowledge learned from observation data to update the displacement field. More precisely, the statFEM allows inference of the system's true state by combining observed data and a stochastic FEM (SFEM) model \cite{keese2003review}. The advantage of this framework is that the SFEM computations can be conducted in an offline design phase. During the updating process in statFEM for sparse online measurements, we only need to condition the state on data and quantify hyperparameters of a Gaussian process model, which describes the model-reality mismatch. \Rev{The effectiveness of statFEM as a valuable instrument for SHM has been previously proposed and put into practice in \cite{febrianto2022digital}. Here, the concept of "statistical digital twin" is developed for a bridge's superstructure, which can merge continuous sensor measurements with FEM model predictions to yield instantaneous and up-to-date statistical projections of strain distribution.  For this purpose, measured strain data are gathered from $108$ Fiber Bragg Grating (FBG) sensors positioned along the primary superstructure of a fully operational railway bridge. It demonstrates that statFEM represents a resilient probabilistic approach for structural health monitoring. In fact}, by adopting a Bayesian formalism, all uncertainties, e.g., material model parameters, measurement errors, and imprecision of the computational model, are quantified. The model discrepancy $\vecd$ is an indispensable part of statFEM. Underlying missing physics, numerical approximations, and other inaccuracies may cause the discrepancy. Arenndt et al. \cite{Arendt2012} use the model discrepancy as a validation criterion to update the model. If the validation shows the model's inaccuracy, we can either gather more observation data or refine the computer model by changing its constitutive model \cite{xiong2009better}.

This paper presents a statFEM representation for solid mechanics with uncertain material parameters. Following Girolami et al. \cite{girolami2021statistical} and the seminal work from Kennedy and O'Hagan  \cite{kennedy2001bayesian}, the data are decomposed into three random components. The first component relies on a stochastic model prediction of the displacements $\vecu$ based on SFEM, which considers the uncertainties in material parameters of different constitutive models. The second component is a model discrepancy term $\vecd$, which reflects the discrepancy between observed data and the response of the calibrated constitutive model. Finally, the third component is a measurement noise term $\vece$, which represents the noise of the sensing device.

The stochastic prior displacement $\vecu$ can be calculated either with classical \textit{Monte Carlo} (MC)  simulation  \cite{hammersley1960monte} or with some alternative approaches that are less computationally demanding. These approaches are divided into two categories: (i) Intrusive and (ii) Non-Intrusive approaches.  The deterministic FEM formulation and corresponding source code need significant modification in intrusive techniques. Widely used intrusive approaches are the Perturbation method \cite{arregui2016practical,kleiber1992stochastic}, Galerkin \textit{Polynomial Chaos} (PC) based on orthogonal polynomials \cite{boyd2001chebyshev,deb2001solution,matthies2005galerkin}, the \textit{Spectral Stochastic Finite Element Method} (SSFEM)  based on the \textit{Karhunen-Lo{\`e}ve expansion} (KLe) for input uncertainty and PC for the stochastic output \cite{ghanem2003stochastic}. On the other hand, the non-intrusive techniques do not require reformulating existing deterministic source code. The non-intrusive approaches are classified into three groups: Regression (least square minimization) \cite{berveiller2005non,berveiller2006stochastic}, Pseudo-Spectral Projection (quadrature based) \cite{reagana2003uncertainty}, and Stochastic Collocation \cite{xiu2005high}. Intrusive methods may require cumbersome manipulations of an existing complex code and are, therefore, not considered here. For this reason, a non-intrusive regression method is chosen in this contribution. \Rev{Although PC has shown to be effective in many applications, it faces limitations when dealing with high-dimensional inputs. This is due to the exponential growth in the number of basis terms and unknown expansion coefficients as the number of input parameters increases, also known as the "curse of dimensionality". To address this issue, sparse quadrature \cite{smolyak1963quadrature,gerstner1998numerical}, \textit{least angle regression} (LAR) \cite{blatman2010adaptive,blatman2011adaptive}, and compressive sensing \cite{doostan2011non,hampton2015compressive} are techniques that can enhance the efficiency and accuracy of the PC approach.}

The contributions of this paper are as follows:
\begin{itemize}
	\item The stochastic forward problem for statFEM based on an uncertain material parameter is modeled with a non-intrusive PC to quantify the uncertain displacement response.
	\item We investigate the concepts of statFEM in the context of linear and non-linear solid mechanics and illustrate the results in $1$D and $2$D examples.
	\item A model selection metric is introduced to assess which material model best explains given observation data.
	\item Inferring displacement fields from sparse online measurement is introduced within the framework of statFEM based on gradient-based estimation of three hyperparameters.
\end{itemize}

The rest of the paper is organized as follows: In \autoref{sec:SMBE}, we define the stochastic boundary value problem. We then introduce the usage of MC and PC expansion to quantify the displacement uncertainty with the help of a simple one-dimensional tension bar. In \autoref{sec:statFEM}, we introduce Gaussian processes, the statFEM method, and the updating procedure based on a Bayesian formalism. A gradient-based algorithm is introduced to identify the hyperparameters. After introducing the fundamental concepts of statFEM, simple one-dimensional results are presented, and the influence of the sensor data on posterior displacement is studied. We then examine the posterior displacement with a linear elastic model and nonlinear observation data. In \autoref{sec:stressField_reconstruction}, we give an extension to $2$D examples to infer the displacement field from sparse online measurement. In \autoref{subsec:materialModelSelection}, we examine which material model best explains the observed data. As an example, an infinite plate with a circular hole with a linear elastic and St.Venant Kirchhoff material model is chosen, and the statistical displacements are presented. Finally, in \autoref{subsec:InferringStressField}, the influence of observation data on the residual of the push-forward stress is discussed.

\section{Stochastic Mechanical Balance Equations}\label{sec:SMBE}
In this section, we review the concepts of stochastic prior based on SFEM. Readers familiar with SFEM may skip this section and continue with \autoref{sec:statFEM}. We begin by defining the stochastic boundary value problem in \autoref{subsec:SBVP}. The section is followed by introducing the PC expansion in \autoref{subsec:PC}. Finally, in \autoref{subsec:1D_SFEM}, we illustrate the concepts of SFEM with a one-dimensional tension bar.

\subsection{Stochastic Boundary Value Problem}\label{subsec:SBVP}
Given a \textit{probability space}  $(\saSp, \sigAl, \prb)$, where $\saSp$ is a \textit{sample space} equipped with the \textit{$\sigma$-algebra} $\sigAl$ and the \textit{probability measure} $\prb$, such that $\prb:\sigAl \rightarrow [0,1]$, and a bounded domain $\Omega \subset \bbR^d$ with $d \in \{1,2,3\}$, the static \textit{stochastic boundary value problem} (SBVP) for deformations consists in seeking a stochastic function $\vecu(\vecX,\theta)$, as in \cite{jahanbin2020stochastic}, such that the following equation holds almost surely:
\begin{equation}
	\begin{cases}
		\grad{\cdot\vecP(\vecX,\theta)} + \vecb(\vecX,\theta)= \vec0,                                         & \text{in} \,\, \Omega                      \\
		\vecu(\vecX,\theta) = \overline{\vecu}(\vecX,\theta),                                                 & \text{on} \,\, \partial\Omega_{\vecu}      \\
		\vecT_{\vecN}(\vecX,\theta) = \vecP(\vecX,\theta)\cdot\vecN = \overline{\vecT}_{\vecN}(\vecX,\theta), & \text{on} \,\, \partial\Omega_{\vecsigma}.
	\end{cases}
	\label{eq:SBVP}
\end{equation}
Here, $\theta \in \saSp$ is an elementary random outcome, $\vecP$ is the \textit{first Piola-Kirchhoff} ($1$.PK) stress tensor,  $\vecb$ denotes the body force, $\overline{\vecu}$ and $\overline{\vecT}_{\vecN}$ are the prescribed displacement and traction on boundaries $\partial\Omega_{\vecu}$ and $\partial\Omega_{\vecsigma}$, respectively, where $\partial\Omega= \partial\Omega_{\vecu} \cup \partial\Omega_{\vecsigma}$ and $\partial\Omega_{\vecu} \cap \partial\Omega_{\vecsigma} =\vecEmpty$. The unit vector normal to boundary $\partial\Omega_{\vecsigma}$ is characterized by $\vecN$. Note that $\vecX$ are the coordinates of the domain $\Omega$ in the reference configuration.

The computation of the stochastic function $\vecu(\vecX,\theta)$ requires to complement \eqref{eq:SBVP} with a material model. \textit{Linear elastic} (LE) and \textit{St. Venant Kirchhoff} (SV) are chosen models in this study. For the linear elastic model, the infinitesimal strain tensor is defined as
\begin{equation}
	\vecepsilon(\vecX,\theta) = \onetwo \big(\grad{\vecu(\vecX,\theta)} +  (\grad{\vecu(\vecX,\theta)})^T\big).
	\label{eq:infinitesimalStrain_linearElastic}
\end{equation}
The stochastic \textit{Cauchy} stress is given as $\vecsigma(\vecX,\theta) = \boldsymbol{\mathcal{D}}(\theta):\vecepsilon(\vecX,\theta)$, where $\boldsymbol{\mathcal{D}}$ is the stochastic fourth-order constitutive tensor. Under the small deformation assumption of the linear elastic model, the first Piola-Kirchhoff and Cauchy stress $\vecsigma$ become identical. For the St. Venant Kirchhoff model, as the simplest hyperelastic model, the stochastic Green-Lagrange strain tensor has the following form
\begin{equation}
	\vecE(\vecX,\theta) = \onetwo \Big(\grad{ \vecu(\vecX,\theta)} +  (\grad{ \vecu(\vecX,\theta)})^T + \grad{ \vecu(\vecX,\theta)} (\grad{ \vecu(\vecX,\theta)})^T\Big),
	\label{eq:greenLangrageStrain_ST}
\end{equation}
and the stochastic \textit{second Piola-Kirchhoff} stress tensor is $\vecS(\vecX,\theta) = \boldsymbol{\mathcal{D}}(\theta):\vecE(\vecX,\theta)$. The $1.$PK is therefore defined as $\vecP = \vecS \vecF^T$, where $\vecF$ is a deformation gradient. The fourth-order constitutive tensor based on uncertain Young's modulus $E(\theta)$ and Poisson's ratio $\nu$ is then given by $\boldsymbol{\mathcal{D}}(\theta) = E(\theta) \cdot \boldsymbol{\mathcal{D}}^*$, where the deterministic $\boldsymbol{\mathcal{D}}^*$ is defined as follows:
\begin{equation}
	\mathcal{D}^*_{ijkl} = \frac{\nu}{(1+\nu)(1-2\nu)}\delta_{ij}\delta_{kl} + \frac{1}{2(1+\nu)} \big(\delta_{il}\delta_{jk} + \delta_{ik}\delta_{jl} \big).
	\label{eq:Deter_stochastic_constitutive}
\end{equation}
Here, $\delta_{ij}$ denotes the Kronecker delta with $i,j,k,l = 1,2,3$. In order to solve the SBVP with FEM, the domain $\Omega$ is discretized into subdomains $\Omega^h$ with finite elements, where the stochastic function $\vecu(\vecX,\theta)$ evaluated at the given set of nodes $\vecX^h = \{\vecX_1^T, \dots, \vecX_{\nnode}^T\}^T$, i.e., $\vecu(\vecX_i,\theta) \approx \vecu^h(\vecX_i,\theta), i=1,\ldots,\nnode$ and  $\vecu^h(\vecX^h,\theta) = \{ \vecu^h(\vecX_1,\theta)^T, \dots, \vecu^h(\vecX_{\nnode},\theta)^T\}^T$ is a vector in $\bbR^{\ngdof}$ with $\ngdof = \nnode \cdot d$, where $\nnode \in \bbN^+$ is the number of nodes. In this study, the only uncertainty in the forward problem is in Young's modulus $E$. The mean vector $\MuMC \in \bbR^{\ngdof}$ and covariance matrix $\CuMC \in \bbR^{\ngdof \times \ngdof}$ of the resulting stochastic displacement can be evaluated either with the well-known MC \cite{schenk2005uncertainty} or with regression as one of the non-intrusive PC approaches.

The mean and covariance matrix of discretized stochastic displacement $\vecu^h(\vecX^h,\theta)$ from $\NMC$ simulations can be approximated with
\begin{equation}
	\MuMC= \frac{1}{\NMC} \sum_{i=1}^{\NMC} \vecu^h(\vecX^h,\theta_i), \quad \text{and} \quad \CuMC = \frac{1}{\NMC-1}\sum_{i=1}^{\NMC} \big(\vecu^h(\vecX^h,\theta_i)-\MuMC \big) \big(\vecu^h(\vecX^h,\theta_i)-\MuMC\big)^T,
	\label{eq:meanCov_MCS}
\end{equation}
where $\bullet_{_\text{MC}}$ indicates that the mean value and covariance matrix are evaluated with Monte Carlo simulation and $\theta_i$ denote i.i.d. realizations. Note that MC is used in this paper as a reference solution for validation.

\subsection{Polynomial Chaos Expansion}\label{subsec:PC}
\Rev{Although the MC simulation is a flexible and robust method, it can often be computationally expensive. Hence, we use the PC technique to represent the stochastic functions. Readers familiar with PC expansion may skip this section and proceed to \autoref{sec:statFEM}.}
The PC expansion method approximates the discretized stochastic function $\vecu^h(\vecX^h,\theta)$ using orthogonal PC bases \cite{ghanem2003stochastic}. It is based on the multiplicative decomposition of a random output into a deterministic part, namely the PC coefficients $\vecu_j$, and a stochastic part $\mPsi_j$ in a separable manner as
\begin{equation}
	\vecu^h(\vecX^h,\theta) =\vecu^h \big(\vecX^h,\vecxi(\theta) \big)= \sum_{j=0}^{\infty} \vecu_j(\vecX^h) \mPsi_j\big(\vecxi(\theta)\big),
	\label{eq:U_PC_inf}
\end{equation}
where, $\vecxi =\{\xi_1, \dots, \xi_M\}$ denotes a set of $M$ independent and identically distributed (i.i.d) standard Gaussian random variables characterizing input uncertainties. We assume that a finite-dimensional germ $\vecxi$ is sufficient to describe the underlying randomness exactly. For practical implementation, the stochastic function $\vecu^h\big(\vecX^h,\vecxi(\theta)\big)$ is approximated by $\vecu_P^h\big(\vecX^h,\vecxi(\theta)\big)$ after truncating it at the $P$-th term, i.e,
\begin{equation}
	\vecu^h \big(\vecX^h,\vecxi(\theta) \big) \approx \vecu_P^h \big(\vecX^h,\vecxi(\theta) \big)= \sum_{j=0}^{P} \vecu_j(\vecX^h) \mPsi_j\big(\vecxi(\theta)\big).
	\label{eq:U_PC}
\end{equation}
The $\vecu_0, \dots, \vecu_P$ are deterministic coefficients, and $\mPsi_0, \dots, \mPsi_P$ are multivariate orthogonal PC bases that depend on the distribution of the random variables $\vecxi$.  In this study, because of the Gaussian distribution of $\vecxi$, we use the probabilistic Hermite polynomials to construct the PC expansion. The multivariate Hermite polynomial $\mPsi_j$ can be shown to be a product of univariate Hermite polynomials $\psi_{\alpha_i^j}$ with a multi-index $(\alpha_1^j,\ldots,\alpha_M^j)$:
\begin{equation}
	\mPsi_j(\vecxi) = \prod_{i=1}^{M} \psi_{\alpha_i^j}\big(\xi_i(\theta)\big).
	\label{eq:multivariatePC}
\end{equation}
To generate the sequence of multi-indices $\alpha_i^j$, we can use either the graded lexicographic ordering method \cite{xiu2010numerical} or the ball-box method  \cite{sudret2000stochastic}. Examples of univariate and multivariate polynomials are provided in \autoref{Appendix:probabilisticHermitie}. Note that the total number of expansion terms $P+1$ in \eqref{eq:U_PC} depends on $M$ and the chosen polynomial order $p$, which can be determined as follows:
\begin{equation}
	P = \frac{(M+p)!}{M!p!} -1.
	\label{eq:POrder}
\end{equation}
The PC coefficients $\vecu_j$ in \eqref{eq:U_PC} can be obtained by the regression method. The regression method, also known as stochastic linear regression, is based on a random sampling of input uncertainty $\vecxi$ and the corresponding propagated output. Given $S \in \bbN^+$ random variables $\vecxi_s \sim \vecxi$ and $\vecu^h \big(\vecX^h,\vecxi_s(\theta) \big)$, $s = 1,\dots,S$, the coefficients $\vecu_j(\vecX^h)$ are computed such that the sum of squared differences between $\vecu_P^h \big(\vecX^h,\vecxi_s(\theta) \big)$ and $\vecu^h \big(\vecX^h,\vecxi_s(\theta) \big)$ is minimized, i.e.
\begin{equation}
	(\vecu_0(\vecX^h),\ldots,\vecu_P(\vecX^h)) = \argmin_{(\vecu_0,\ldots,\vecu_P)} \frac{1}{S} \sum_{s=1}^{S} \Big[ \vecu^h \big(\vecX^h,\vecxi_s(\theta) \big)- \sum_{j=0}^{P} \vecu_j(\vecX^h) \mPsi_j\big(\vecxi_s(\theta) \big)\Big]^2
	\label{eq:PC_minimization}
\end{equation}
which leads to the following expression
\begin{equation}
	\vecc = (\vecA^T\vecA)^{-1}\vecA^T\vecz
	\label{eq:LS}
\end{equation}
where,
\begin{equation}
	\vecA =
	\begin{bmatrix}
		\mPsi_0\big(\vecxi_0\big) & \cdots & \mPsi_P\big(\vecxi_0\big) \\
		\mPsi_0\big(\vecxi_1\big) & \cdots & \mPsi_P\big(\vecxi_1\big) \\
		\vdots                    & \cdots & \vdots                    \\
		\mPsi_0\big(\vecxi_S\big) & \cdots & \mPsi_P\big(\vecxi_S\big)
	\end{bmatrix}, \,\,\,
	\vecc =
	\begin{bmatrix}
		\vecu_0^T(\vecX^h) \\
		\vecu_1^T(\vecX^h) \\
		\vdots             \\
		\vecu_P^T(\vecX^h)
	\end{bmatrix}, \,\,\,
	\vecz =
	\begin{bmatrix}
		\vecu^h(\vecX^h,\vecxi_0)^T \\
		\vecu^h(\vecX^h,\vecxi_1)^T \\
		\vdots                      \\
		\vecu^h(\vecX^h,\vecxi_S)^T
	\end{bmatrix}.
	\label{eq:PointCollocationMatrix}
\end{equation}
The suitable number of $S$ in \eqref{eq:PC_minimization} depends on $\vecu^h \big(\vecX^h,\vecxi(\theta) \big)$, the order of polynomial bases, and also the required accuracy. Interested readers can refer to \cite{Berveiller2005,stigler1971optimal} for more information. Consequently, the response $\vecu^h \big(\vecX^h,\vecxi(\theta) \big)$ can be obtained with substituting the coefficients $\vecu_0,\dots,\vecu_P$ in \eqref{eq:U_PC}. Therefore, the mean vector $\MuPC  \in \bbR^{\ngdof}$ and covariance matrix $\CuPC \in \bbR^{\ngdof \times \ngdof}$ of the response vector are obtained as follows:
\begin{equation}
	\MuPC = \vecu_0(\vecX^h), \quad \text{and} \quad \CuPC= \sum_{j=1}^{P} \langle\, \mPsi_j^2\big(\vecxi\big) \,\rangle  \vecu_j(\vecX^h) \vecu_j^T(\vecX^h).
	\label{eq:meanCovPC}
\end{equation}
Here, $\langle\,  \bullet   \,\rangle$ is the expectation over $\vecxi$, and $\bullet_{_\text{PC}}$ indicates that the mean value and covariance matrix are evaluated with the PC expansion. In this contribution, the only random variable in the forward problem is an uncertain Young's modulus, i.e., $M = 1$. Therefore, we rewrite \eqref{eq:U_PC} for the univariate case as follows:
\begin{equation}
	\vecu^h \big(\vecX^h,\xi(\theta) \big) \approx \vecu_P^h \big(\vecX^h,\xi(\theta) \big)= \sum_{j=0}^{P} \vecu_j(\vecX^h) \psi_j\big(\xi(\theta)\big).
	\label{eq:U_PC_univariate}
\end{equation}
We have to expand the uncertain material parameter $E$ in PC format. Because Young's modulus must be strictly non-negative, it is represented by a lognormal distribution with a mean $\mu_E$ and standard deviation $\sigma_E$, $E \sim \mathcal{LN}\big(\mu_E\,,\, \sigma_E^2\big)$, and is defined, as in \cite{ang2007probability,cao2017probabilistic}, as follows:
\begin{equation}
	E \big(\xi(\theta) \big) = \exp{\mu_{\kappa}+\sigma_{\kappa}\xi(\theta)}.
	\label{eq:LognormalYoungModulus}
\end{equation}
Here, $\mu_{\kappa}$ and $\sigma_{\kappa}$ are the mean and standard deviation of $\kappa=\ln{E}$, respectively. The $\mu_{\kappa}$ and $\sigma_{\kappa}$ of a Gaussian distributed $\kappa$ can be obtained from the following relations:
\begin{equation}
	\mu_{\kappa} = \ln{\frac{\mu_E^2}{\sqrt[]{\mu_E^2+\sigma_E^2}}}, \quad  \sigma_{\kappa}^2 = \ln{1+\frac{\sigma_E^2}{\mu_E^2}}.
	\label{eq:LognormalTransform}
\end{equation}
Aiming to represent the lognormal distribution by a Hermite polynomial, we need more terms in the PC expansion. The PC expansion of a lognormal random Young's modulus with Hermite polynomials can therefore be presented after truncating it at the $P_E$-th term as in \cite{xiu2010numerical,ghanem1999nonlinear}:
\begin{equation}
	E \big(\xi(\theta) \big) \approx E_{P_E} \big(\xi(\theta) \big)= \sum_{i=0}^{P_E} \mu_E \frac{\sigma_\kappa^i}{i !} \psi_i\big(\xi(\theta)\big).
	\label{eq:material_PC}
\end{equation}
The fourth-order stochastic constitutive tensor is then given by $\boldsymbol{\mathcal{D}}\big(\xi(\theta) \big) = E_{P_E} \big(\xi(\theta) \big) \boldsymbol{\mathcal{D}}^*$, where $\boldsymbol{\mathcal{D}}^*$ is the deterministic part of the stochastic constitutive tensor defined in \eqref{eq:Deter_stochastic_constitutive}. The matrix-vector notation of this formulation is shown in \autoref{Appendix:MVConstitutive}.

\subsection{One-dimensional SFEM Example} \label{subsec:1D_SFEM}
\begin{figure}[!htp]
	\centering
	\includegraphics{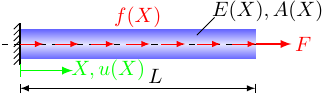}
	\caption{One-dimensional bar under tension.}
	\label{fig:1D_bar_geometry}
\end{figure}

In this section, we briefly illustrate the concepts of FEM and SFEM using a one-dimensional linear elastic bar under tension as depicted in \autoref{fig:1D_bar_geometry}. Although this part is standard and well-known, we need this example and the following results to demonstrate the results of the statFEM.

The strong deterministic form of this problem is
\begin{equation}
	\begin{split}
		-\frac{d}{dX} &\bigg[E(X)A(X) \frac{d u(X)}{dX}\bigg] - f(X) = 0, \quad X \in [0,L] \\
		u(X)&\bigg|_{X=0} = 0, \,\,\,\,\,\, \frac{d u(X)}{dX}\bigg|_{X=L} = \frac{F}{E(X)A(X)} \bigg|_{X=L},
	\end{split}
	\label{eq:1D_bar_PDE}
\end{equation}
where $L$ is the length of the bar, $E(X)$ Young's modulus, $A(X)$ the cross-sectional area, $u(X)$ the displacement, $F$ the concentrated load at the tip of the bar and $f(X)$ a continuously distributed line load as unit force per unit length. Moreover, the two Dirichlet and Neumann boundary conditions are also given in \eqref{eq:1D_bar_PDE}. After formulating the weak form of \eqref{eq:1D_bar_PDE}, and solving the well-known FEM system of equations \Rev{with $49$ elements}, we obtain the discretized displacement $\vecu^h(\vecX^h) \Rev{\in \bbR^{50 \times 1}}$.
\begin{figure}[!ht]
	\centering
	\subfloat[]{
		\centering
		\includegraphics{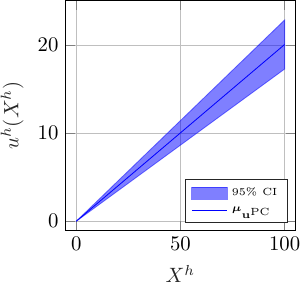}
		\label{subfig:1D_bar_mean_CI_MC}}
	\hspace*{0.7cm}
	\subfloat[]{
		\centering
		\includegraphics{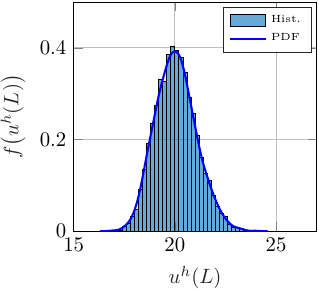}
		\label{subfig:1D_bar_HisPDF}}
	\\
	\vspace*{-0.4cm}
	\subfloat[]{
		\centering
		\includegraphics{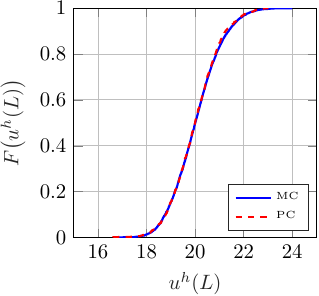}
		\label{subfig:1D_bar_CDF_MCPC}}
	\hspace*{0.7cm}
	\subfloat[]{
		\centering
		\includegraphics{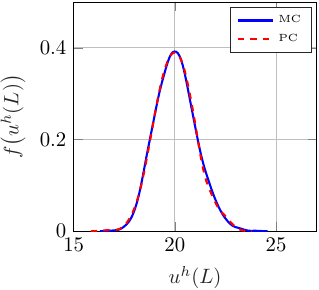}
		\label{subfig:1D_bar_PDF_MCPC}}
	\caption{\textbf{Prior Displacement of $1$D Tension Bar}: (a) Mean displacement and its $95\%$ credible interval (CI); (b) PDF of displacement at tip of the bar; (c) Comparison of CDF of displacement at tip of the bar; and, (d) Comparison of PDF of displacement at tip of the bar.}
	\label{fig:1D_bar_uncertain1}
\end{figure}
As an special case, where $L = 100 \,\si{\mm}$, $E(X) = 200 \,\si{\GPa}$, $A(X) = 20 \, \si{\mm}$, $f(X) = 0 \, \si{kN \, \mm^{-1}}$ and $F=800 \, \si{kN}$ are constants, the homogeneous analytical displacement $u_{\text{ho}}(X)$ is:
\begin{equation}
	u_{\text{ho}}(X) = \frac{1}{EA}(FX + fLX - \frac{f}{2}X^2) =  \frac{F}{EA}X\,.
	\label{eq:1D_bar_u_analytic}
\end{equation}

To formulate the SBVP, we consider the input uncertainty in the material parameter $E \sim \mathcal{LN}\big(\mu_E=200 \,\si{\GPa}\,,\, \sigma_E^2=10^2 \,\si{\GPa}^2 \big)$. Note that in all examples of this contribution, we chose $P_E = 4$. \autoref{fig:1D_bar_uncertain1} demonstrates the results of stochastic displacement based on MC\Rev{, $\MuMC \in \bbR^{50 \times 1} $,} with $\NMC=1000$ against PC\Rev{, $\MuPC \in \bbR^{50 \times 1}$,} with $P=5$. \autoref{subfig:1D_bar_HisPDF} shows the comparison of histogram and \textit{probability distribution function} (PDF) at the tip of the bar. \autoref{subfig:1D_bar_CDF_MCPC} and \autoref{subfig:1D_bar_PDF_MCPC} depict a reasonable agreement between PDFs and \textit{cumulative distribution function} (CDF) between MC and PC at the tip of the bar. \Rev{It is noteworthy that that we obtained the stochastic displacement by solving the forward problem only $S = 2P = 10$ times using PC which is a significant reduction compared to the $\NMC=1000$ times required by the MC method.  To provide a better understanding of the computational complexity, let $n_{\vecu}=49$ be the number of unknown degrees of freedom. Typically, the FEM generates a linear system of equations of size $n_{\vecu} \times n_{\vecu}$. Solving this system of equations has a computational complexity of $\calO(n_u^s)$, where $s$ is typically 1, 2 or 3 depending on the linear solver used. Therefore, with PC, the computation requires $\calO(2P n_u^s)$ to obtain the stochastic displacement.}

\section{Statistical Finite Element Method}\label{sec:statFEM}
In order to present the statFEM approach, first, Gaussian process models are defined in \autoref{subsec:DGP}. We then introduce the statistical generating model, the data conditioning procedure, and the hyperparameter estimation method in \autoref{subsec:statFEM_theory}. Finally, in \autoref{subsec:statFEM_1D_results}, we show the posterior results of the one-dimensional tension bar for linear and nonlinear observation data.
\subsection{Gaussian Process Models}\label{subsec:DGP}
Given the probability space, $(\saSp, \sigAl, \prb)$, a stochastic process $H(\vecX,\theta)$ is defined as a collection of random variables that describe a random quantity at coordinate $\vecX \in \Omega$ on a bounded domain $\Omega$. The process $H(\vecX,\theta)$ is said to be a \textit{Gaussian process} (GP), if for every set of points $\vecX_i \in \Omega, i =1,\dots,n$, the $n$-variate distribution of $\{ H(\vecX_1,\theta), \dots, H(\vecX_n,\theta)\}^T$ is jointly Gaussian, see e.g. \cite{shi2011gaussian}. We can fully specify a GP by its mean function $\mu_H:\Omega \rightarrow \bbR$ and covariance function $c_H: \Omega \times \Omega \rightarrow \bbR$. The covariance function can be expressed as $c_H(\vecX,\vecX') = \sigma_H(\vecX) \cdot \sigma_H(\vecX') \cdot \rho_H(\vecX,\vecX')$, where $\sigma_H:\Omega \rightarrow \mathbb{R}^+_0$ is a standard deviation function and $\rho_H: \Omega \times \Omega \rightarrow [-1,1]$ is a correlation coefficient function. If $H$ is a Gaussian process, we write
\begin{equation}
	H(\vecX) \isGP \big(\mu_H(\vecX), c_H(\vecX,\vecX') \big).
	\label{eq:GP_definition}
\end{equation}
In this context, the covariance function $c_H(\vecX,\vecX')$ is also referred to as kernel or kernel function interchangeably. By definition, the kernel is bounded and symmetric. Moreover, we work with strictly positive definite kernels, i.e., the matrix obtained by evaluating the kernel $\vecC_H$ at $\vecX_i$ with $i =1,\ldots,n$ is positive definite. A widely used kernel function is the exponential quadratic kernel
\begin{equation}
	c_H(\vecX,\vecX') = \sigma_H^2 \exp{-\frac{||\vecX-\vecX'||^2}{2l_H^2}},
	\label{eq:SE_Kernel}
\end{equation}
where we assumed $  \sigma_H(\vecX) = \sigma_H(\vecX') =  \sigma_H$.

\subsection{Formulation of StatFEM}\label{subsec:statFEM_theory}
\begin{figure}[!htb]
	\centering
	\includegraphics{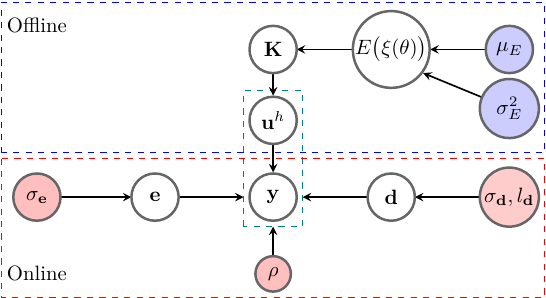}
	\caption{\textbf{Graphical Illustration of StatFEM}: The colored circles are the parameters, either known or to be identified, and the white circles are either observed or derived. The blue circles are the parameters associated with the forward problem, which are assumed to be known. The red circles are the hyperparameters, which can be estimated based on observations $\yObs$. This figure is adapted from \cite{girolami2021statistical}.}
	\label{fig:graphicalstatFEM}
\end{figure}

The core part of statFEM is the so-called statistical generating model. This model is based on the observed data, a computational forward model, and all possible uncertainties, which may contain additional unknown parameters. The graphical illustration of statFEM, adapted from \cite{girolami2021statistical}, can be seen in \autoref{fig:graphicalstatFEM}. Here, the pair $\big(\mu_E$, $\sigma_E^2\big)$ represents the uncertainty of Young's modulus, which is propagated to the stiffness matrix $\vecK$.

Suppose we are collecting a single observation and store in a vector $\yObs \in \bbR^{\ngsen}$  or multiple observations in a matrix $\YObs \in \bbR^{\ngsen \times \nrep }$ from a physical system on a set of sensor points $\xsen \in \bbR^{\ngsen}$, with $\nsen \in \bbN^+$ referring to the number of sensors, $\nrep \in \bbN^{+}$ the number of readings from a sensor, $d$ the observation dimension, and $\ngsen = \nsen \cdot d$. Note that the observed data $\YObs$ may differ from $\zRel$, the true data on sensors, because of sensor measurement errors, which are contained in a vector $\vece \in  \bbR^{\ngsen}$. Following the framework proposed in \cite{girolami2021statistical,kennedy2001bayesian}, there is a model-reality mismatch, which also has to be considered. The mismatch is, in fact, the discrepancy between the simulated data from the computer model and the physical response, which is represented with a regression parameter $\rho \in  \bbR^+$ and a model uncertainty vector $\vecd \in \bbR^{\ngsen}$. The statistical generating model can now be defined as
\begin{equation}
	\yObs = \zRel + \vece =  \rho \vecH \vecu^h(\vecX^h, \theta) + \vecd + \vece,
	\label{eq:statGenerModel}
\end{equation}
where $\vecu^h(\vecX^h, \theta) \in \bbR^{\ngdof}$ consists of the nodal displacments such that $\nnode \gg \nsen$, and $\vecH \in \bbR^{\ngsen \times \ngdof}$ the projection matrix. The regression parameter $\rho$ is used to weigh the evidence provided by the simulator term $\vecH \vecu^h(\vecX^h, \theta)$ relative to the model discrepancy.

Following the statFEM framework in \cite{girolami2021statistical}, we assume that both model discrepancy $\vecd$ and measurement error $\vece$ can be modeled independently by GPs. Moreover, we model each component of the vectorial discrepancy as an independent GP, i.e.
\begin{equation}
	\vecd_i(\vecX) \isGP \big(\mu_{\vecd_i}(\vecX), c_{\vecd_i}(\vecX,\vecX') \big), \ \ i=1,\ldots,d,
	\label{eq:d_GP}
\end{equation}
with kernel function
\begin{equation}
	c_{\vecd_i}(\vecX,\vecX') = \sigma_{\vecd_i}^2 \exp{-\frac{||\vecX-\vecX'||^2}{2l_{\vecd_i}^2}}.
	\label{eq:SE_Kernel_d}
\end{equation}
We then model each entry of vector $\vecd_i$ and $\vece_i$, evaluated at sensor locations $\xsen$, as an $\nsen$-dimensional Gaussian vector
\begin{equation}
	\vecd_i(\xsen) \sim \prb (\vecd_i(\xsen) | \sigma_{\vecd_i}, l_{\vecd_i}) = \mathcal{N}\bigg(\vec0, \vecC_{\vecd_i}\bigg),
	\label{eq:discrepancyGP}
\end{equation}
and
\begin{equation}
	\vece_i(\xsen)\sim \mathcal{N}\bigg(\vec0, \vecC_{\vece_i}\bigg).
	\label{eq:errorGP}
\end{equation}
Repeated observations are modeled with independent Gaussians. Note that the covariance matrix $\vecC_{\vecd_i} \in \bbR^{\nsen \times \nsen}$ is obtained through evaluating the kernel function at $\xsen$ and noise covariance matrix $\vecC_{\vece_i} \in \bbR^{\nsen \times \nsen}$ is defined with $\vecC_{\vece_i} = \sigma_{\vece_i}^2 \vecI$.  The hyperparameters $\sigd=\sigma_{\vecd_i}$, $\ld=l_{\vecd_i}$, and $\rho$, which present the model discrepancy, are collected in a vector
\begin{equation}
	\vecw : = \big( \rho,\sigd,\ld \big).
	\label{eq:hyperparameterVec}
\end{equation}

According to the Bayesian formalism, the posterior density of the non-intrusive SFEM solution given observation data and the hyperparameters is formulated by
\begin{equation}
	\prb(\vecu^h | \YObs, \vecw) = \frac{\prb(\YObs | \vecu^h, \vecw)\prb(\vecu^h)}{\prb(\YObs |  \vecw )}.
	\label{eq:statFEM_Bayesian}
\end{equation}
Similar to \cite{girolami2021statistical}, we approximate the density of $\prb(\vecu^h)$ with the multivariate Gaussian density, i.e., $\prb(\vecu^h)= \mathcal{N}\big(\MuPC, \CuPC \big)$, where the PC expansion is used to approximate the unknown second order moments of the displacement. The density of $\vecu$ is usually not Gaussian, and the approximation is only valid for small $\sigma_E$. Because the readings from sensors are independent of each other, the data likelihood is given by
\begin{equation}
	\prb(\YObs | \vecu^h, \vecw) = \prb(\yObsOne | \vecu^h, \vecw)\prb(\yObsTwo | \vecu^h, \vecw) \cdots\prb(\yObsNrep | \vecu^h, \vecw) = \prod_{i=1}^{\nrep}\prb(\YObsi | \vecu^h, \vecw),
	\label{eq:likelihoodJoint}
\end{equation}
and the marginal likelihood by $\prb(\YObs|  \vecw ) = \prod_{i=1}^{\nrep}\prb(\YObsi|  \vecw )$ with
\begin{equation}
	\prb(\YObsi|  \vecw ) = \int \prb(\YObsi |\vecu^h, \vecw)\prb(\vecu^h) \d{\vecu^h} = \mathcal{N}\bigg(\rho \vecH \MuPC, \CD + \CE + \rho^2 \vecH \CuPC \vecH^T \bigg)
	\label{eq:marginalLikelihoodSigle}
\end{equation}
with $\vecC_{\vecd},\vecC_{\vece} \in \bbR^{\ngsen \times \ngsen}$.

The posterior density is then a multivariate Gaussian and is given by
\begin{equation}
	\prb(\vecu^h |\yObsOne, \yObsTwo, \dots, \yObsNrep, \vecw ) = \prb(\vecu^h |  \YObs, \vecw) = \mathcal{N}\bigg(\MuPCY, \CuPCY \bigg),
	\label{eq:posteriorMultivariateGaussian}
\end{equation}
where the mean vector $\MuyPCY \in \bbR^{\ngdof}$ and the covariance matrix $\CuPCY \in \bbR^{\ngdof \times \ngdof}$ are defined as
\begin{equation}
	\MuPCY = \CuPCY \bigg(\vecH^T (\CD + \CE)^{-1} \sum_{i=1}^{\nrep} \YObsi  + (\rho \CuPC)^{-1} \MuPC  \bigg)
	\label{eq:postU}
\end{equation}
and
\begin{equation}
	\CuPCY = \bigg(\nrep \vecH^T (\CD + \CE)^{-1} \vecH + (\rho^2 \CuPC)^{-1} \bigg)^{-1}.
	\label{eq:postCU}
\end{equation}
The two results in \eqref{eq:postU} and \eqref{eq:postCU} can be interpreted as a correction of the SFEM results given the observation data. Another viewpoint considers \eqref{eq:postU} as a sensor data regression with a physically motivated regression function. In the Bayesian formalism, these results are the posterior of the mean displacement and covariance with the prior inputs from \eqref{eq:meanCovPC}. With these results at hand, the probability of the \Rev{"}unobserved\Rev{"} true system response is given by

\begin{equation}
	\prb (\zRel | \YObs) = \mathcal{N}\bigg(\vecH \MuPCY, \vecH \CuPCY \vecH^T + \CD \bigg)=\mathcal{N}\bigg(\Muz, \CZ \bigg).
	\label{eq:postYreal}
\end{equation}
\Rev{We note that $\zRel$ represents the true displacement on the sensor locations. According to \eqref{eq:statGenerModel}, $\zRel$ consists of projected FEM outcomes, $\rho \vecH \vecu^h(\vecX^h, \theta)$, scaled by an unknown regression parameter $\rho$, and an unknown model-reality mismatch, $\vecd$. Here, "unobserved" signifies that the true displacement $\zRel$ on sensor devices cannot be measured or observed directly. Instead, it is inferred from the measurements at the sensor locations and the finite element model. To elaborate further, \eqref{eq:postU} and \eqref{eq:postCU} update the mean value and covariance of the displacement on the FEM nodes, whereas \eqref{eq:postYreal} transforms this update to the sensor location.  It can be deduced from  \eqref{eq:postYreal} that the mean value of the true displacement on the sensor $\Muz$ is the projection of $\MuPCY$ on sensor locations with the help of $\vecH$. Moreover, the covariance matrix of the true displacement on the sensors $\CZ$ is a projection of $\CuPCY$ and an additional term $\CD$. The covariance matrix $\CD$ considers the discrepancy between the model and the reality, which arises from the presence of the unobserved term $\vecd$ in \eqref{eq:statGenerModel}. In this work, the projection matrix $\vecH$ serves as a transformative tool, projecting the displacement of FEM nodes onto the sensor location via the utilization of shape functions and the coordinates of the sensor location within each element. However, in general, $\vecH$ can exhibit a higher degree of flexibility based on the characteristics of the experimental data. For instance, if the observation data are strains, the computation of $\vecH$ becomes more complicated, relying on the derivative of shape functions.}

The hyperparameters in $\vecw$, which describe the model discrepancy, are unknown beforehand. Therefore, the estimation procedure of hyperparameters $\vecw^*  = \big( \rho^*,\sigd^*,\ld^* \big)$ is explained separately in \autoref{Appendix:hyperparameterEstimation}. As can be seen  \Rev{in \autoref{fig:graphicalstatFEM}}, statFEM offers a framework to infer the displacement field from observation data by solving the forward problem only once\Rev{, which constitutes the offline phase of the statFEM method}. This is extra helpful for real-time reconstruction of the displacement field. \Rev{The online phase of the statFEM involves applying Bayesian techniques to update the prior displacement given observation data in a probabilistic framework.}  \Rev{This phase consists of two parts. The first part involves solving the optimization problem \eqref{eq:minMarginalDerivative} to identify the hyperparameters, while the second part involves obtaining $\MuPCY$ and $\CuPCY$ from \eqref{eq:postU} and \eqref{eq:postCU}. The identification step can be seen as a minimization problem. To demonstrate the computational complexity of this step, let $k$ donate the number of iterations required by the quasi-newton algorithm to converge to a solution. We need a Cholesky decomposition of a dense positive-definite matrix $\vecvarSigma = \CD + \CE + \rho^2 \vecH \CuPC \vecH^T$, providing both numerical stability and computational efficiency \cite{golub2013matrix}, for every iteration, which can be performed for $\calO(\ngsen^3)$. The inverse and determinant of $\vecvarSigma$ is needed, as seen in \eqref{eq:thetaFun}, which requires  $\calO(\ngsen^s)$. The total computational complexty of first step is then $\calO\big(k(\ngsen^3+ \ngsen^s)\big) = \calO\big(k\ngsen^3\big)$. The second part needs an inverse of a dense matrix $\CD + \CE$ at the cost of $\calO\big(\ngsen^3\big)$, which is similar to the first step. Note that the inverse of large dense $\CuPC$ is performed at the cost of $\calO(n_\vecu^3)$. In online regime,} for a new set of sparse measurements, we only need to identify the three hyperparameters $\rho^*,\sigd^*,\ld^*$ and update the existing stochastic prior results with \eqref{eq:postU} and \eqref{eq:postCU}. \Rev{According to \cite{duffin2022low}, the issue of computational scalability presents a challenge for applying statFEM to high-dimensional problems commonly encountered in physical and industrial contexts. Therefore, a low-rank extended Kalman filter algorithm for statFEM is proposed to address this challenge, which has demonstrated high scalability, particularly for nonlinear, time-dependent problems.}

\Rev{To analyze the impact of observation data richness on the updating process, it is necessary to understand the correlation between $\CD + \CE$ and $\CuPC$ with $\CuPCY$. To clarify this relationship, we initially assume $\nsen = 1$ and $\rho=1$. It should be noted that \eqref{eq:postCU} is obtained by using the Sherman-Morrison-Woodbury identity, as in \cite{girolami2021statistical}, of the following equation:
	\begin{equation}
		\begin{split}
			\CuPCY & =  \CuPC - \underbrace{\CuPC \vecH^T\bigg(\vecH \CuPC \vecH^T + \CD + \CE \bigg)^{-1}}_\text{\pmb{\calK}} \vecH \CuPC \\
			\CuPCY & = \CuPC - \pmb{\calK} \vecH \CuPC = (\vecI - \pmb{\calK}\vecH) \CuPC.
		\end{split}
	\end{equation}
	In this context, $\pmb{\calK}$ represents the Kalman gain. There are two possible extreme scenarios to consider: Firstly, when $(\CD + \CE) \rightarrow \infty$, indicating that $\CD + \CE$ is much larger than $\CuPC$, then $\pmb{\calK}$ approaches zero, resulting in $\CuPCY \rightarrow \CuPC$. This implies that the extremely large value of $\CD + \CE$ corresponds to poor-informative observation data that does not significantly impact the updating process. Consequently, no shrinkage occurs in $\CuPC$. Secondly, when $(\CD + \CE) \rightarrow 0$, representing that $\CD + \CE$ is significantly smaller than $\CuPC$, the Kalman gain $\pmb{\calK}$ approaches $\vecH^{-1}$, leading to $\CuPCY \rightarrow (\CD + \CE) \rightarrow 0$. The minimal value of $\CD + \CE$ implies that the observation data are rich-informative and greatly influence the updating process.}

\subsection{Simple One-dimensional StatFEM Results}\label{subsec:statFEM_1D_results}
In \autoref{subsubsec:statFEM_sensor_influence}, we apply the statFEM framework on the tension bar for linear generated observation data. Then, in \autoref{subsubsec:statFEM_nonlinearData}, we present the posterior displacement with nonlinear generated observation data. The difference between this section and the contribution in \cite{girolami2021statistical} is the usage of PC-based stochastic prior and the estimation of hyperparameters based on a gradient-based optimizer.

\subsubsection{Influence of Sensors on Posterior Results}\label{subsubsec:statFEM_sensor_influence}
In order to illustrate the concepts of statFEM, we use the results of the linear elastic example in \autoref{subsec:1D_SFEM} as prior information in the statFEM framework. The observation data are generated synthetically based on the analytical displacement solution in \eqref{eq:1D_bar_u_analytic} with predefined hyperparameters. Although these hyperparameters are unknown in real cases, we predefined them to generate the synthetic data and examine the correctness of \autoref{Appendix:hyperparameterEstimation}. The predefined hyperparameters are:
\begin{equation}
	\vecw = \big( \rho = 0.7,\,\, \sigd = 0.9,\,\,\ld = 2.0 \big), \quad \sigma_\vece^2 = 0.004.
	\label{eq:hyperparameterNoise}
\end{equation}
The numbers of sensors are $\nsen= \{11, 33\}$ and the numbers of readings from sensors are chosen $\nrep = \{1, 10,100\}$. The identified hyperparameters based on \autoref{Appendix:hyperparameterEstimation} are summarized in \autoref{tab:identified}.
\begin{table}[!htb]
	\centering
	\begin{tabular}{cccc}
		\toprule
		$\rho^*$   & $\nrep=1$  & $\nrep=10$ & $\nrep=100$ \\ \midrule
		$\nsen=11$ & $0.811901$ & $0.767034$ & $0.731313$  \\
		$\nsen=33$ & $0.760251$ & $0.748731$ & $0.691733$  \\
		\bottomrule
	\end{tabular} \\ \vspace*{0.2cm}
	\centering
	\begin{tabular}{cccc}
		\toprule
		$\sigd^*$  & $\nrep=1$  & $\nrep=10$ & $\nrep=100$ \\ \midrule
		$\nsen=11$ & $0.673814$ & $0.700081$ & $0.797473$  \\
		$\nsen=33$ & $0.695891$ & $0.755548$ & $0.895953$  \\
		\bottomrule
	\end{tabular} \\ \vspace*{0.2cm}
	\centering
	\begin{tabular}{cccc}
		\toprule
		$\ld^*$    & $\nrep=1$  & $\nrep=10$ & $\nrep=100$ \\ \midrule
		$\nsen=11$ & $1.514279$ & $1.554831$ & $1.741418$  \\
		$\nsen=33$ & $1.413163$ & $1.544271$ & $1.958134$  \\
		\bottomrule
	\end{tabular}
	\caption{The identified hyperparameters.\label{tab:identified}}
\end{table}

\begin{figure}[!htb]
	\centering
	\hspace*{-0.5cm}
	\subfloat[$\nrep=1$, $\nsen = 11$]{
		\centering
		\includegraphics{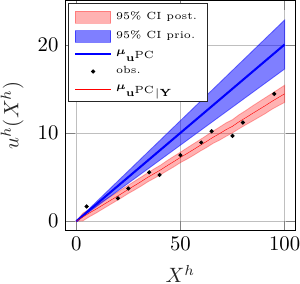}
		\label{subfig:1D_bar_post1_11}}
	\subfloat[$\nrep=10$, $\nsen = 11$]{
		\centering
		\includegraphics{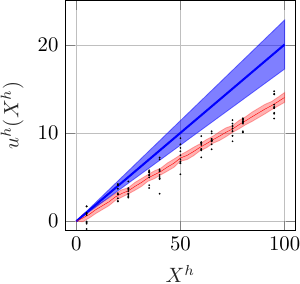}
		\label{subfig:1D_bar_post10_11}}
	\subfloat[$\nrep=100$, $\nsen = 11$]{
		\centering
		\includegraphics{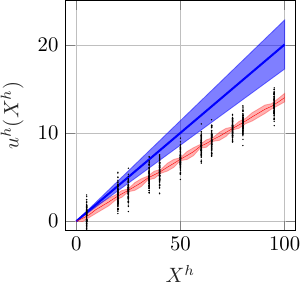}
		\label{subfig:1D_bar_post100_11}}
	\\
	\vspace*{-0.2cm}
	\hspace*{-0.5cm}
	\subfloat[$\nrep=1$, $\nsen = 33$]{
		\centering
		\includegraphics{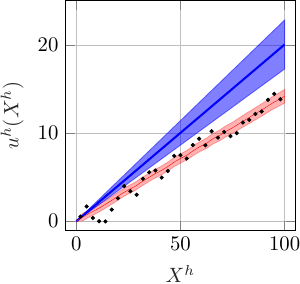}
		\label{subfig:1D_bar_post33_1}}
	\subfloat[$\nrep=10$, $\nsen = 33$]{
		\centering
		\includegraphics{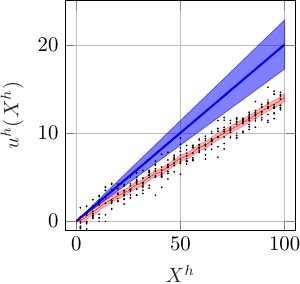}
		\label{subfig:1D_bar_post33_10}}
	\subfloat[$\nrep=100$, $\nsen = 33$]{
		\centering
		\includegraphics{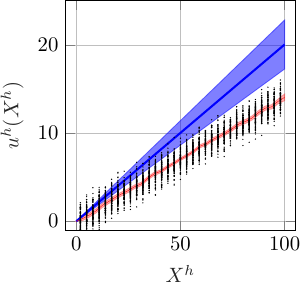}
		\label{subfig:1D_bar_post33_100}}
	\caption{\textbf{Posterior Displacement of $1$D Tension Bar}: The blue lines represented $\MuPC$ and shaded blue areas denote the $95\%$ CI of $\MuPC$. The red lines represented $\MuPCY$ and shaded red areas denote the $95\%$ CI of $\MuPCY$. The black dots are observation data. The style of representation is adapted from \cite{girolami2021statistical}.}
	\label{fig:1D_bar_posterior}
\end{figure}

As seen in \autoref{fig:1D_bar_posterior}, even with small numbers of $\nrep$ and $\nsen$, the solution $\MuPCY \Rev{\in \bbR^{50 \times 1}}$ lies on the observation data. The uncertainty in the posterior solution can be reduced in terms of mean value and covariance with the increasing number of sensors $\nsen$ and more readings $\nrep$. As seen in \autoref{subfig:1D_bar_post33_100}, the $\MuPCY$ almost overlaps the mean of the observation data with a narrow CI.

\begin{figure}[!htb]
	\centering
	\subfloat[]{
		\centering
		\includegraphics{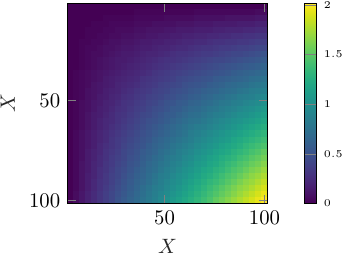}
		\label{subfig:1D_bar_prior_Cu_heatmap}}
	\hspace*{0.7cm}
	\subfloat[$\nrep = 100$, $\nsen=33$]{
		\centering
		\includegraphics{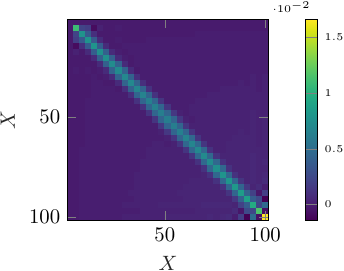}
		\label{subfig:1D_bar_poster_Cu_heatmap}}
	\caption{\textbf{Heatmap of Covariance}: (a) Covariance matrix of prior displacement given uncertain Young's modulus $\CuPC$; (b) Covariance matrix of posterior displacement given observation data $\CuPCY$.}
	\label{fig:1D_bar_covariance_heatmap}
\end{figure}

\begin{figure}[H]
	\centering
	\hspace*{-0.5cm}
	\subfloat[$\nrep=1$, $\nsen = 11$]{
		\centering
		\includegraphics{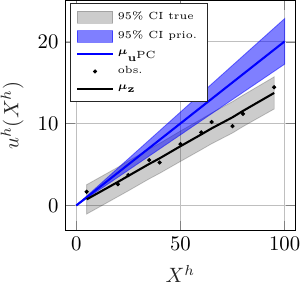}
		\label{subfig:1D_bar_true1_11}}
	\subfloat[$\nrep=10$, $\nsen = 11$]{
		\centering
		\includegraphics{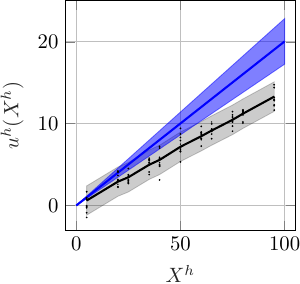}
		\label{subfig:1D_bar_true10_11}}
	\subfloat[$\nrep=100$, $\nsen = 11$]{
		\centering
		\includegraphics{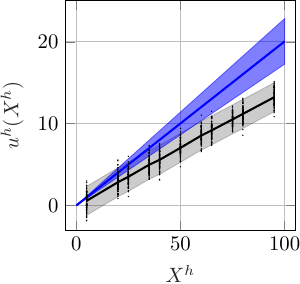}
		\label{subfig:1D_bar_true100_11}}
	\\
	\vspace*{-0.2cm}
	\hspace*{-0.5cm}
	\subfloat[$\nrep=1$, $\nsen = 33$]{
		\centering
		\includegraphics{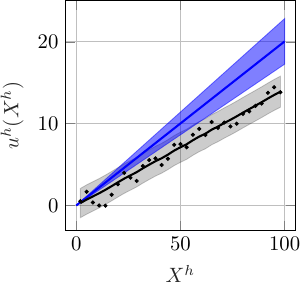}
		\label{subfig:1D_bar_true1_33}}
	\subfloat[$\nrep=10$, $\nsen = 33$]{
		\centering
		\includegraphics{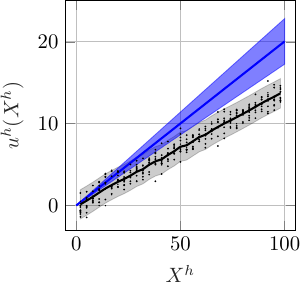}
		\label{subfig:1D_bar_true10_33}}
	\subfloat[$\nrep=100$, $\nsen = 33$]{
		\centering
		\includegraphics{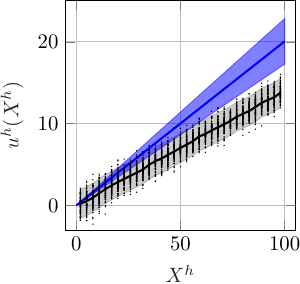}
		\label{subfig:1D_bar_true100_33}}
	\caption{\textbf{True Displacement on Sensor Locations of $1$D Tension Bar}: The blue lines represented $\MuPC$ and shaded blue areas denote the $95\%$ CI of $\MuPC$. The black lines represented $\zRel$ and shaded black areas denote the $95\%$ CI of $\zRel$. The black dots are observation data on sensor locations. The style of representation is adapted from \cite{girolami2021statistical}.}

	\label{fig:1D_bar_trueResponse}
\end{figure}

\autoref{fig:1D_bar_covariance_heatmap} illustrates the influence of observation data on a reduction of the covariance. The behavior of the true system response on sensor locations\Rev{, $\zRel \in \bbR^{\nsen \times 1}$,} is shown in \autoref{fig:1D_bar_trueResponse}. The shaded gray area represents the model-reality mismatch, reflected in the statistical generating model $\vecd$. Quantifying model-reality error based on observation data is one of the major results of statFEM.

\subsubsection{Linear Model with Nonlinear Data}\label{subsubsec:statFEM_nonlinearData}
\begin{figure}[!htb]
	\centering
	\subfloat[]{
		\centering
		\includegraphics{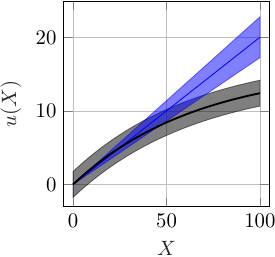}
		\label{subfig:1D_nonbar_analytical_trueResponse}}
	\subfloat[]{
		\centering
		\includegraphics{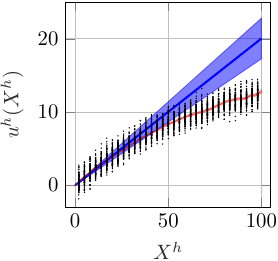}
		\label{subfig:1D_nonbar_postu}}
	\subfloat[]{
		\centering
		\includegraphics{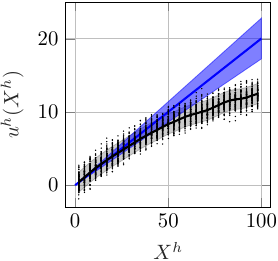}
		\label{subfig:1D_nonbar_postz}}
	\caption{\textbf{Posterior and True Displacement of $1$D Tension Bar with Nonlinear Observation Data}: (a) The blue line shows $\MuPC$, and the blue shaded area is the $95\percent$ of CI. The black line shows the true analytical displacement with a $95\percent$ CI. (b) Posterior of displacement given observation data: The black dots are the generated observation data based on \eqref{eq:1D_bar_u_analytic_inhomo} and \eqref{eq:hyperparameterNoise_inhomo}. The red lines represented $\MuPCY$ and the shaded red area denotes its $95\%$ CI; (c) True displacement: The black lines represented $\zRel$ and the shaded black area denotes its $95\%$ CI.}
	\label{fig:1D_nonbar_uncertain}
\end{figure}

This section examines the results of statFEM in light of nonlinear observation data. For this purpose, we assume a heterogeneous Young's modulus, given by
\begin{equation}
	E(X) = E_0 \exp{\beta X}, \quad \text{with} \, \, \beta = 0.015 \,\si{\mm}^{-1} \,, E_0 = 200 \,\si{\GPa}.
	\label{eq:inhomo_E}
\end{equation}
The inhomogeneous analytical displacement is given by the following equation
\begin{equation}
	u_{\text{inho}}(X) = \frac{F}{E_0 \beta A}\Big(1-\exp{-\beta X}\Big).
	\label{eq:1D_bar_u_analytic_inhomo}
\end{equation}
Here, $\beta$ is a given constant and if $\beta \rightarrow 0$, then $u_{\text{inho}} \rightarrow u_{\text{ho}}$. Consequently, we generate the observation data based on the nonlinear function \eqref{eq:1D_bar_u_analytic_inhomo} with uncertain material parameter $E_0 \sim \mathcal{LN}\big(\mu_{E_0}=200 \,\si{\GPa}\, , \, \sigma_{E_0}^2=10^2 \,\si{\GPa}^2 \big)$ and the following predefined hyperparameters.
\begin{equation}
	\vecw = \big( \rho = 1.2,\,\, \sigd = 0.9,\,\,\ld = 4.0 \big), \quad \sigma_\vece^2 = \Rev{0.004}.
	\label{eq:hyperparameterNoise_inhomo}
\end{equation}
Given the predefined hyperparameters, we can obtain the analytical true system response as shown in \autoref{subfig:1D_nonbar_analytical_trueResponse}.

\begin{figure}[!htb]
	\centering
	\subfloat[$\nrep = 100$, $\nsen=33$]{
		\centering
		\includegraphics{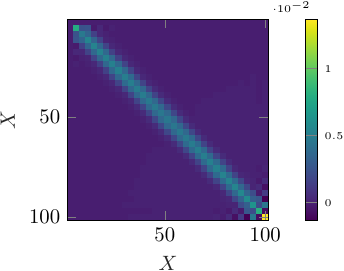}
		\label{subfig:1D_nonbar_poster_Cu}}
	\hspace*{0.7cm}
	\subfloat[$\nrep = 100$, $\nsen=33$]{
		\centering
		\includegraphics{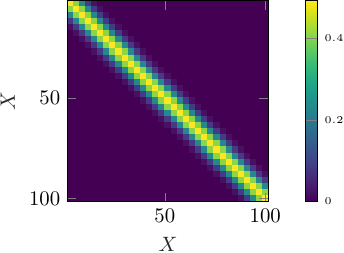}
		\label{subfig:1D_nonbar_poster_Cz}}
	\caption{\textbf{Heatmap of Covariance}: (a) $\CuPCY$ given non-linear observation data. (b) The covariance matrix of true displacement on sensor locations.}
	\label{fig:1D_nonbar_covariance}
\end{figure}

As it can be seen in \autoref{fig:1D_nonbar_uncertain}, with $\nsen = 33$ and $\nrep = 100$, the $\MuPCY \Rev{\in \bbR^{50 \times 1}}$ shows a good agreement with the observation data, and the $95\%$ CI covers a small region. In \autoref{fig:1D_nonbar_covariance}, the posterior covariance $\CuPCY \Rev{\in \bbR^{50 \times 50}}$ becomes smaller in light of observation data compared to the prior covariance matrix $\CuPC \Rev{\in \bbR^{50 \times 50}}$ shown in \autoref{subfig:1D_bar_prior_Cu_heatmap}. \autoref{subfig:1D_nonbar_poster_Cz} illustrates the covariance of $\zRel \Rev{\in \bbR^{\nsen \times 1}}$. The wider diagonal represents the model-reality mismatch, which is being quantified with the help of identified hyperparameters.

\begin{figure}[!htb]
	\centering
	\subfloat[$\sigma_\vece^2=0.04$]{
		\centering
		\includegraphics{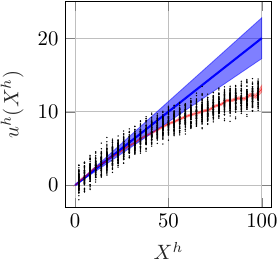}
		\label{subfig:1D_nonbar_Noise1}}
	\subfloat[$\sigma_\vece^2=0.4$]{
		\centering
		\includegraphics{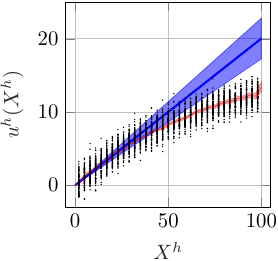}
		\label{subfig:1D_nonbar_Noise2}}
	\subfloat[$\sigma_\vece^2=4$]{
		\centering
		\includegraphics{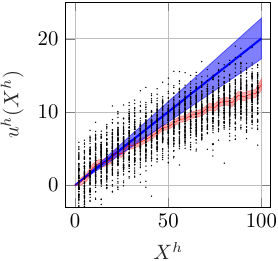}
		\label{subfig:1D_nonbar_Noise3}}
	\caption{\Rev{\textbf{Posterior Displacement for Different Noise Levels}: (a) Posterior of displacement given observation data generated with $\sigma_\vece^2=0.04$; (b) Posterior of displacement given observation data generated with $\sigma_\vece^2=0.4$; (c) Posterior of displacement given observation data generated with $\sigma_\vece^2=4$.}}
	\label{fig:1D_nonbar_Noise}
\end{figure}

An exciting conclusion of this section is the excellent agreement of $\MuPCY$ based on the linear elastic SFEM model as prior information for non-linear observation data. We can conclude that if enough observation data exist, even a simple material model can update the displacement.
\Rev{It should be noted that, the chosen noise in \eqref{eq:hyperparameterNoise_inhomo} is intentionally small to focus on identifying the hyperparameters such as $\sigd$, $\ld$, and $\rho$,  which arise from the model-reality mismatch. In practice, the noise produced by measurement devices is typically unknown. Therefore, an additional hyperparameter $\sigma_\vece$ is required to minimize the equation \eqref{eq:minMarginal}. To examine the effect of high noise, we conducted computations with varying noise levels $\vecsigma_\vece^2 = \{0.04,\, 0.4,\, 4.0 \}^T$ and then evaluate the influence of noise on the resulting posterior displacement. The results can be seen in \autoref{fig:1D_nonbar_Noise}. It can be concluded that higher noise levels lead to increased uncertainty in the updated displacement. In this case, the identified hyperparameters are not changing significantly with the varying noise level.}
\section{Material Model Selection and Stress Inference with StatFEM}\label{sec:stressField_reconstruction}
Another important aspect of statFEM is to clarify which prior material model explains observation data better. In \autoref{subsec:materialModelSelection}, we examine the inferred displacement field for prior LE and SV material models in light of sparse observation data. Then, in \autoref{subsec:InferringStressField} we examine the residual in the balance of linear momentum for prior and posterior stresses. We label the posterior stress as a posterior push-forward stress field.

\subsection{Material Model Selection}\label{subsec:materialModelSelection}
\begin{figure}[!htb]
	\centering
	\subfloat[]{
		\centering
		\includegraphics{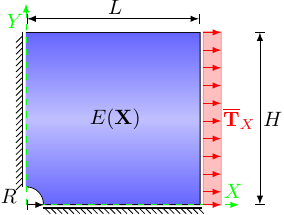}
		\label{subfig:2D_ip_Geometry}}
	\hspace*{2cm}
	\subfloat[]{
		\centering
		\includegraphics[height=0.2\textwidth]{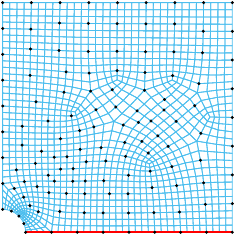}
		\label{subfig:2D_ip_meshing_sensors}}
	\caption{\textbf{Setup of an Infinite Plate with Hole}: (a) Geometry, loading, and boundary condition. (b) The blue lines show geometry meshing, the black dots are the sensor locations, and the red line indicates the reference line.}
	\label{fig:2D_ip_meshing_sensors}
\end{figure}

In this section, we examine the influence of the prior material model on the posterior results, and we demonstrate which material model best explains observation data. We first assume that the material is a rubber-like material. The geometry of the plate is shown in \autoref{subfig:2D_ip_Geometry}, where $R = 0.4$ is the radius, $L = 4$ is the length, $H = L$ is the height, and the applied load is a traction $\overline{\vecT}_X = 100\,\si{\MPa}$ in $X$-direction. The corresponding mesh and the sensor locations are illustrated in \autoref{subfig:2D_ip_meshing_sensors}. The \Rev{number of elements is $1200$,} the number of nodes is $\nnode = 1273$, and the number of sensors is $\nsen = 127$. \Rev{The standard Lagrange linear shape functions were employed for the elements, and integration was performed using $4$ Gauss points.} The material parameters are Young's modulus $E = 200\,\si{\GPa}$\Rev{,} and Poisson's ratio $\nu = 0.25$.

\begin{figure}[!htb]
	\centering
	\subfloat[$\MuxPC$ with LE]{
		\centering
		\includegraphics{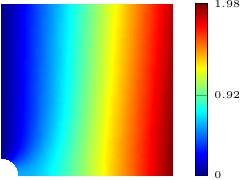}
		\label{subfig:2D_ip_FEM_mean_ux_PC_LE}}
	\hspace*{0.5cm}
	\subfloat[$\STDuxPC$ with LE]{
		\centering
		\includegraphics{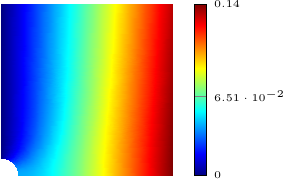}
		\label{subfig:2D_ip_FEM_std_ux_PC_LE}}
	\hspace*{0.5cm}
	\subfloat[$\STDuxMC$ with LE]{
		\centering
		\includegraphics{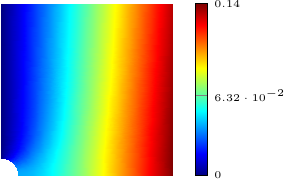}
		\label{subfig:2D_ip_FEM_std_ux_MC_LE}}
	\\
	\hspace*{-0.6cm}
	\subfloat[$\MuxPC$ with SV]{
		\centering
		\includegraphics{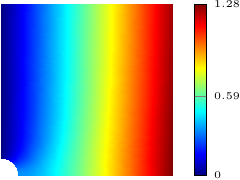}
		\label{subfig:2D_ip_FEM_mean_ux_PC_SV}}
	\hspace*{0.6cm}
	\subfloat[$\STDuxPC$ with SV]{
		\centering
		\includegraphics{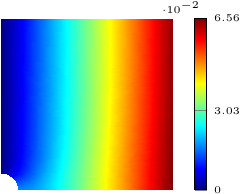}
		\label{subfig:2D_ip_FEM_std_ux_PC_SV}}
	\hspace*{1.3cm}
	\subfloat[$\STDuxMC$ with SV]{
		\centering
		\includegraphics{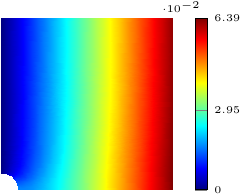}
		\label{subfig:2D_ip_FEM_std_ux_MC_SV}}
	\caption{\textbf{Prior Results}: (a) Prior mean of horizontal displacement field $\MuxPC$ from PC with LE model; (b) Prior standard deviation of horizontal displacement field $\STDuxPC$ from PC with LE model; (c) Prior standard deviation of horizontal displacement field $\STDuxMC$ from MC with LE model; (d) Prior mean of horizontal displacement field $\MuxPC$ from PC with SV model; (e) Prior standard deviation of horizontal displacement field $\STDuxPC$ from PC with SV model; (f) Prior standard deviation of horizontal displacement field $\STDuxMC$ from MC with SV model.}
	\label{fig:2D_ip_FEM_moments_ux_PC}
\end{figure}

\begin{figure}[!htb]
	\centering
	\subfloat[]{
		\includegraphics{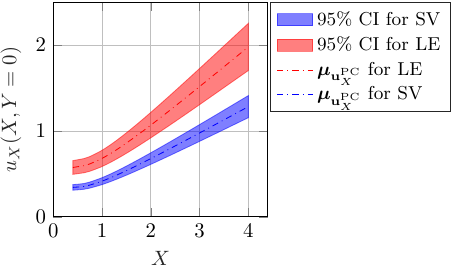}
		\label{subfig:2D_ip_SFEM_u_refrenceLine}}
	\hspace*{0.3cm}
	\subfloat[]{
		\includegraphics{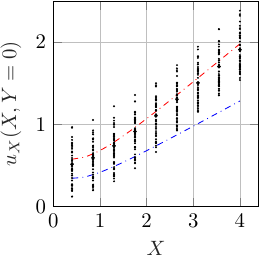}
		\label{subfig:2D_ip_u_y_n50}}
	\caption{\textbf{Prior Horizontal Displacement on the Reference Line}: (a) The dashed red line is the mean of the prior horizontal displacement from the LE material model on the reference line with its 95$\percent$ CI. The dashed blue line is the mean of the prior horizontal displacement from the SV material model on the reference line with a 95$\percent$ CI. (b) The black dots are $\nrep = 50$ observation data of sensors on the reference line and the bold black dots are the mean value of observations.}
	\label{fig:2D_ip_refrenceLine1}
\end{figure}

We then simulate the SFEM solution of the problem with two material models, namely the linear elastic (LE) and the nonlinear hyperelastic St. Venant Kirchhoff (SV) material model. The stochastic displacement field is evaluated with uncertain Young's modulus $E(\vecX) \sim \mathcal{LN}\big(\mu_E=200 \,\si{\GPa}\, , \, \sigma_E^2=20^2  \,\si{\GPa}^2\big)$. In this step, we use the regression method to estimate the PC coefficients of the displacement field with $P = 9$ and $S = 2P = 18$ regression points. Note that because of the nonlinearity involved in the SV model displacement-strain relations, we use the iterative Newton-Raphson method. \Rev{The outputs of SFEM are the mean vector and covariance matrix of the displacement field, denoted as $\MuPC \in \bbR^{2546 \times 1}$ and $\CuPC \in \bbR^{2546 \times 2546}$, respectively, which are computed using \eqref{eq:meanCovPC}.} The contour plots of the mean horizontal displacement field \Rev{$\MuxPC \in \bbR^{1273 \times 1}$} and its standard deviation \Rev{$\STDuxPC \in \bbR^{1273 \times 1}$}in $X$-direction for both LE and SV models are presented in \autoref{fig:2D_ip_FEM_moments_ux_PC}. Moreover, the stochastic displacement field is calculated with PC, and the accuracy is verified with an MC simulation with $\NMC = 10,000$.
In order to compare the results of FEM, SFEM, and statFEM more easily, we choose the horizontal displacement along $Y=0$ as a reference line; see the red line in \autoref{subfig:2D_ip_meshing_sensors}. The comparison of SFEM results on the reference line for LE and SV models are presented in \autoref{subfig:2D_ip_SFEM_u_refrenceLine}. In the next step, we generate synthetic observation displacement data on sensor locations based on the displacement solution obtained with the SV model and the predefined hyperparameters given as
\begin{equation}
	\vecw = \big( \rho = 1.5,\,\, \sigd = 0.2,\,\,\ld = 2.0 \big), \quad \sigma_\vece^2 = 0.0004.
	\label{eq:hyperparameterNoise_2d_ip}
\end{equation}
The choice of the SV model to generate the observation data is because of our initial assumption. We assumed that the material is rubber-like; logically, a hyperelastic material model, e.g., the SV model\Rev{, $\mathcal{M}_{\text{SV}}$}, is more suitable for generating noisy mismatched observation data than the LE model\Rev{, $\mathcal{M}_{\text{LE}}$}. In \autoref{subfig:2D_ip_u_y_n50}, the generated observation data on the sensors located on the reference line is shown.
In reality, we only have access to displacement observation data on sensors. Therefore, we identify the hyperparameters for two cases; for LE and SV material models, and the results are summarized in \autoref{tab:identified2}.
\begin{table}[htb]
	\centering
	\begin{tabular}{cccc}
		\toprule
		          & predefined & LE       & SV       \\ \midrule
		$\rho^*$  & $1.5$      & $0.8455$ & $1.5013$ \\
		$\sigd^*$ & $0.2$      & $0.2405$ & $0.2007$ \\
		$\ld^*$   & $2.0$      & $2.0867$ & $1.9972$ \\
		\bottomrule
	\end{tabular} 	\caption{Prescribed and identified hyperparameters for the example in \autoref{subsec:materialModelSelection}.\label{tab:identified2}}
\end{table}
The different results for LE and SV material models reveal an important conclusion. The choice of the prior material model influences the identified hyperparameters and, subsequently, the inferred displacements. A simple comparison between the predefined and identified hyperparameters shows that the SV material model explains observation data better, as expected.

\begin{figure}[!htb]
	\centering
	\subfloat[Posterior results from LE]{
		\includegraphics{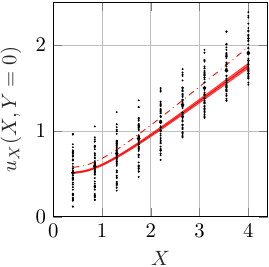}
		\label{subfig:2D_ip_statFEM_refrenceLine_LE}}
	\hspace*{1.0cm}
	\subfloat[Posterior results from SV]{
		\includegraphics{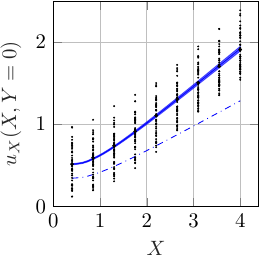}
		\label{subfig:2D_ip_statFEM_refrenceLine_SV}}
	\caption{\textbf{Posterior Horizontal Displacement on the Reference Line}: (a) The dashed red line is the mean of the prior horizontal displacement, and the red line represents the posterior horizontal displacement with its 95$\percent$ CI from the LE material model; (b) The dashed blue line is the mean of the prior horizontal displacement, and the blue line represents the posterior horizontal displacement with its a 95$\percent$ CI from the SV material model.}
	\label{fig:2D_ip_statFEM}
\end{figure}

However, in practice, we only have access to displacement observation data and not predefined hyperparameters, which makes the previous comparison impossible. We, therefore, need a metric for model comparison.  In this study, we \Rev{first} utilize the simpler \textit{Root Mean Square Error} (RMSE) metric, \Rev{as suggestd in \cite{duffin2022statistical},} to select the better model after inferring the displacement field
\begin{equation}
	\text{RMSE} = \frac{1}{\nrep}\sum_{i=1}^{\nrep} \sqrt{\frac{||\Muz - \YObsi||^2}{\nsen}}.
	\label{eq:RMSE}
\end{equation}
The statFEM results on the reference line based on LE and SV material models are shown in \autoref{subfig:2D_ip_statFEM_refrenceLine_LE} and \autoref{subfig:2D_ip_statFEM_refrenceLine_SV}, respectively. As can be seen, the mean posterior displacement from the SV model \Rev{$\mathcal{M}_{\text{SV}}$} is in better agreement with observation data than the LE model \Rev{$\mathcal{M}_{\text{LE}}$}. This can be shown in \autoref{tab:RMSE} based on RMSE for each prior model.
\begin{table}[htb]
	\centering
	\begin{tabular}{ccc}
		\toprule
		Model & \Rev{$\ModelLE$} & \Rev{$\ModelSV$} \\ \midrule
		RMSE  & $0.3558$         & $0.2922$         \\
		\bottomrule
	\end{tabular} 	\caption{The RMSE metric for LE and SV.\label{tab:RMSE}}
\end{table}
The RMSE metric demonstrates that the posterior mean error from the SV model is smaller than the LE model, and the SV model best explains the data. \Rev{Another well-known metric is the \textit{Bayes Factor} ($BF$), which compares the posterior probabilities of competing models, e.g., LE and SV, as suggested in \cite{girolami2021statistical,mackay2003information,murphy2012machine}. In order to evaluate the commitment of two models, namely \Rev{$\ModelLE$} and \Rev{$\ModelSV$}, in describing the observation data \YObs, we calculate the ratio of their posterior probabilities using Bayes' formula with:
	\begin{equation}
		\frac{\prb( \ModelSV | \YObs)}{\prb( \ModelLE | \YObs)}= \frac{\prb(\YObs | \ModelSV)\prb(\ModelSV)}{\prb(\YObs | \ModelLE)\prb(\ModelLE)},
		\label{eq:statFEM_Bayesian_Ratio}
	\end{equation}
	and the Bayes factor of the SV model over the LE model, $\BFSVLE$, is then defined as
	\begin{equation}
		\BFSVLE = \frac{\prb( \ModelSV | \YObs)\prb(\ModelLE)}{\prb( \ModelLE | \YObs)\prb(\ModelSV)}= \frac{\prb(\YObs | \ModelSV)}{\prb(\YObs | \ModelLE)}.
		\label{eq:statFEM_Bayesian_BF0}
	\end{equation}
	Under the assumption of equal prior information for the two models, we can set the prior probabilities as $\prb(\ModelSV) = \prb(\ModelLE) = 0.5$. If $\BFSVLE > 1$, we prefer the SV model over the LE model. Conversely, we prefer the LE model if $\BFSVLE < 1$. It is worth noting that $\prb(\YObs | \ModelLE)$ and $\prb(\YObs | \ModelSV)$ correspond to the marginal likelihood, as defined in \eqref{eq:statFEM_Bayesian}, which is conditioned on the hyperparameters identified with the LE and SV models, respectively. Its closed form is given in \eqref{eq:marginalLikelihoodFull} and the logarithmic form in \eqref{eq:thetaFun}. The computed value of $\BFSVLE = 1.075 > 1$ indicates a preference for the SV model. However, since the factor is only slightly greater than $1$, there is no strong preference for the SV model over the LE model.
}

\begin{figure}[!htb]
	\centering
	\subfloat[$\MuxPCY$ with LE]{
		\centering
		\includegraphics{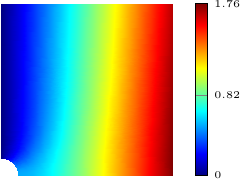}
		\label{subfig:2D_ip_statFEM_mean_ux_PC_LE}}
	\hspace*{0.5cm}
	\subfloat[$\SuxPCY$ with LE]{
		\centering
		\includegraphics{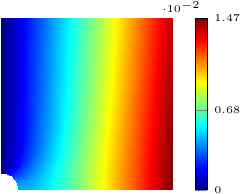}
		\label{subfig:2D_ip_statFEM_std_ux_PC_LE}}
	\\
	\subfloat[$\MuxPCY$ with SV]{
		\centering
		\includegraphics{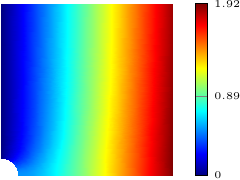}
		\label{subfig:2D_ip_statFEM_mean_ux_PC_SV}}
	\hspace*{0.5cm}
	\subfloat[$\SuxPCY$ with SV]{
		\centering
		\includegraphics{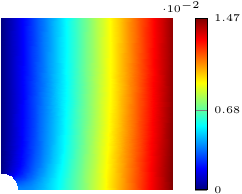}
		\label{subfig:2D_ip_statFEM_std_ux_PC_SV}}
	\caption{\textbf{Posterior Results}: (a) Posterior mean of horizontal displacement field $\MuxPCY$ with LE model; (b) Posterior standard deviation of horizontal displacement field $\SuxPCY$ with LE model; (c) Posterior mean of horizontal displacement field $\MuxPCY$ with SV model; (d) Posterior standard deviation of horizontal displacement field $\SuxPCY$ with SV model.}
	\label{fig:2D_ip_statFEM_moments_ux_PC}
\end{figure}
The results of posterior displacement are shown in \autoref{fig:2D_ip_statFEM_moments_ux_PC}. It is clear, that  $\MuxPCY$ in \autoref{subfig:2D_ip_statFEM_mean_ux_PC_LE} shows smaller values based on the LE model compared to $\MuxPC$ in  \autoref{subfig:2D_ip_FEM_mean_ux_PC_LE} but the $\MuxPCY$ in \autoref{subfig:2D_ip_statFEM_mean_ux_PC_SV} shows bigger displacement compared to $\MuxPC$ in \autoref{subfig:2D_ip_FEM_mean_ux_PC_SV} based on the SV model. The comparison of \autoref{subfig:2D_ip_FEM_std_ux_PC_LE} with \autoref{subfig:2D_ip_statFEM_std_ux_PC_LE} as well as the comparison of \autoref{subfig:2D_ip_FEM_std_ux_PC_SV} with \autoref{subfig:2D_ip_statFEM_std_ux_PC_SV} show a reduction of the standard deviation of the displacement field for both models in light of observation data.

Note that while the observation data are collected online, the solution of the stochastic forward problem can be conducted prior to the observation in an offline stage. Furthermore, for every new set of observation displacement, we estimate the hyperparameters to update the inferred displacement field. Therefore, this method is a promising tool for digital twinning and structural health monitoring.

\subsection{Stress Inference}\label{subsec:InferringStressField}
We updated the displacement field in the previous section based on sparse observation displacement data. Then we utilized the RMSE \Rev{and Bayes factor} metric\Rev{s} to assess how well a model explains the observation data. \Rev{These metrics demonstrated that $\ModelSV$ is more accurate than $\ModelLE$ to update the displacement field. However, two questions must be addressed to infer the stresses: 1) Which model, e.g., $\ModelSV$ or $\ModelLE$, should generate the posterior push-forward stresses? and 2) Is an update of the stochastic parameter, e.g., the Young's modulus, necessary, or is the given $E = 200\,\si{\GPa}$ sufficiently accurate? To determine which model is best for generating the push-forward stresses, we compared the residuals in the prior and posterior for both models, focusing on} the equilibrium equation in \eqref{eq:SBVP}. \Rev{ The prior residual should be almost zero for both models. The model with smaller posterior residual indicates that it is more accurate for generating the push-forward stress. However, to address the second question regarding whether an update of Young's modulus is necessary, a framework is needed to update both the displacement field and Young's modulus simultaneously. This is an area of our future research.}

In this section, we construct the posterior push-forward stresses without model-reality mismatch. We show that the equilibrium is almost fulfilled \Rev{for $\ModelSV$} if the only contamination in observation data is a small noise. For this purpose, we generate the synthetic displacement data on the sensor locations with the following predefined hyperparameters
\begin{equation}
	\vecw = \big( \rho = 1.0,\,\, \sigd = 0,\,\,\ld = 2.0 \big), \quad \sigma_\vece^2 = 1e^{-10}.
	\label{eq:hyperparameterNoise_2d_ip_withoutModelReality}
\end{equation}
\Rev{We intentionally generated the observation data with minimal noise and no model-reality mismatch to highlight the residual difference between the prior and posterior more clearly.} The predefined hyperparameters lead to a zero matrix for $\vecC_\vecd$. Therefore, the only covariance matrix inhered from observation data in \eqref{eq:postU} and \eqref{eq:postCU} is $\vecC_\vece$. Alternatively, the statistical generating model in \eqref{eq:statGenerModel} is reformulated as $\yObs = \vecH \vecu^h(\vecX^h, \theta) + \vece$, with $\vecu^h(\vecX^h, \theta)$  derived from SFEM with the SV model.

As mentioned before, in reality, we only have access to displacement observation data on sensors. Therefore, we identify the hyperparameters for two cases; for LE and SV material models, and the results are summarized in \autoref{tab:identified3}. It shows that the identified hyperparameters based on the SV model are better in agreement with predefined hyperparameters.
\begin{table}[htb]
	\centering
	\begin{tabular}{cccc}
		\toprule
		          & predefined & LE         & SV         \\ \midrule
		$\rho^*$  & $1.0$      & $0.999821$ & $1.0$      \\
		$\sigd^*$ & $0.0$      & $0.999951$ & $0.000001$ \\
		$\ld^*$   & $2.0$      & $0.999886$ & $2.0$      \\
		\bottomrule
	\end{tabular} 	\caption{Prescribed and identified hyperparameters for the example in \autoref{subsec:InferringStressField}.\label{tab:identified3}}
\end{table}

\begin{figure}[!htb]
	\centering
	\subfloat[$\MsigmaxxPC$ from LE]{
		\centering
		\includegraphics{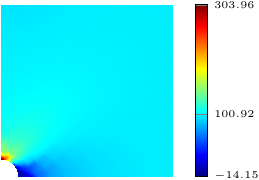}
		\label{subfig:2d_ip_sfem_Cauchy_nomismatch_LE}}
	\hspace*{1.0cm}
	\subfloat[$\MsigmaxxPCY$ from LE]{
		\centering
		\includegraphics{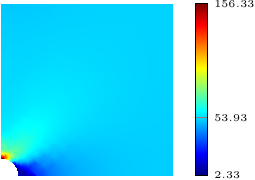}
		\label{subfig:2d_ip_statfem_Cauchy_nomismatch_LE}} \\
	\subfloat[$\MeanPiolaPC$ from SV]{
		\centering
		\includegraphics{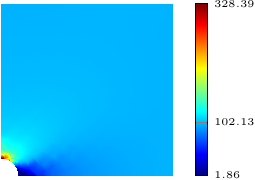}
		\label{subfig:2d_ip_sfem_firstPK_nomismatch_SV}}
	\hspace*{1.0cm}
	\subfloat[$\MeanPiolaPCY$ from SV]{
		\centering
		\includegraphics{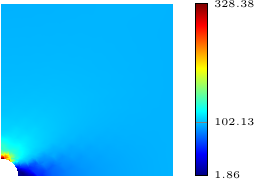}
		\label{subfig:2d_ip_statfem_firstPK_nomismatch_SV}}
	\caption{\textbf{Stress Results}: (a) Mean Cauchy stress from LE in $X$-direction $\MsigmaxxPC$; (b) Posterior push-forward Cauchy stress from LE in $X$-direction $\MsigmaxxPCY$; (c) Mean $1$.PK stress from SV in $X$-direction $\MeanPiolaPC$; (d) Posterior push-forward $1$.PK stress from SV in $X$-direction $\MeanPiolaPCY$.}
	\label{fig:2d_ip_mean_sfem_statfem_Stresses_nomismatch}
\end{figure}

Similar to the previous section, we update the displacement fields for both models. Given $\MuPCY$ on the FE nodes, derivative of shape functions, and $\mu_E$, we obtain an approximation to the Cauchy stress for the LE model $\MsigmaPCY$ and the $1$.PK stress for the SV model $\MPiolaPCY$, which we label as posterior push-forward stresses. This procedure is similar to deterministic FEM. The mean prior stresses are already calculated with MC with $\NMC = 10,000$ in \autoref{subsec:materialModelSelection}. The mean prior stresses and posterior push-forward stresses in $X$-direction are represented in \autoref{fig:2d_ip_mean_sfem_statfem_Stresses_nomismatch}.

\begin{figure}[!htb]
	\centering
	\subfloat[Prior residual with LE]{
		\centering
		\includegraphics{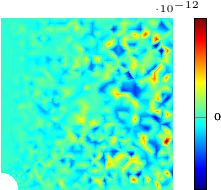}
		\label{subfig:2d_ip_sfem_residual_nomismatch_LE}}
	\hspace*{1.1cm}
	\subfloat[Posterior residual with LE]{
		\centering
		\includegraphics{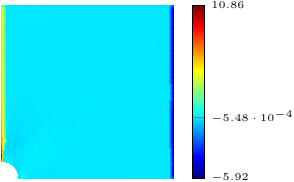}
		\label{subfig:2d_ip_statfem_residual_nomismatch_LE}} \\
	\hspace*{-1.3cm}
	\subfloat[Prior residual with SV]{
		\centering
		\includegraphics{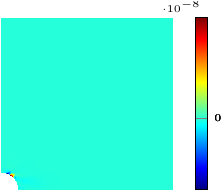}
		\label{subfig:2d_ip_sfem_residual_nomismatch_SV}}
	\hspace*{0.9cm}
	\subfloat[Posterior residual with SV]{
		\centering
		\includegraphics{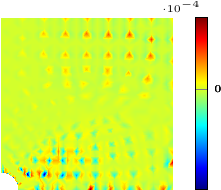}
		\label{subfig:2d_ip_statfem_residual_nomismatch_SV}}
	\caption{\textbf{Residual in the balance of linear momentum}: (a) Residual for mean prior stress in $X$-direction from LE; (b) Residual for mean posterior push-forward stress in $X$-direction from LE; (c) Residual for mean prior stress in $X$-direction from SV; (d) Residual for mean posterior push-forward stress in $X$-direction from SV.}
	\label{fig:2d_ip_mean_sfem_statfem_residual_nomismatch}
\end{figure}

In order to answer which push-forward stress, $\MsigmaxxPCY$ or $\MeanPiolaPCY$, is the correct inferred stress, we have to examine the residual of the balance of linear momentum (\ref{eq:SBVP}) for the prior and posterior stresses.  The residual in $X$-direction is shown in \autoref{fig:2d_ip_mean_sfem_statfem_residual_nomismatch} for LE and SV material models. Comparing \autoref{subfig:2d_ip_sfem_residual_nomismatch_LE} with \autoref{subfig:2d_ip_statfem_residual_nomismatch_LE} shows that posterior push-forward stresses with the LE model violate the equilibrium dramatically. A comparison of \autoref{subfig:2d_ip_sfem_residual_nomismatch_SV} with \autoref{subfig:2d_ip_statfem_residual_nomismatch_SV} demonstrates that the SV model is a better material model to infer the stress field. In this case, the difference between the divergence of prior stress and the divergence of posterior push-forward stress is less significant. This difference originates from the existing noise in observation data. The influence of noise is not recognizable if we compare $\MeanPiolaPC$ with $\MeanPiolaPCY$, because the chosen noise is small. Adding the model-reality mismatch to the observation data makes this difference bigger, and constructing the posterior push-forward stresses become more challenging. If a deviation in the material parameter causes the model-reality mismatch, the material parameter has to be corrected to ensure an accurate stress prediction. This is one of our topics for future work.

\section{Conclusion}\label{sec:conclusions}
This paper presents a mechanics-oriented version of the statFEM approach, which can be used to infer displacement fields from sparse displacement data. The statFEM method interpolates data based on stochastic finite element models based on physical principles. We have employed the well-established non-intrusive Polynomial Chaos method to obtain a model-based prior, where the uncertainties originate in unknown material parameters. The non-intrusive approach allows for handling complex nonlinear mechanics problems with existing deterministic codes. The solution to this stochastic problem is then conditioned on data with Bayesian principles. During Bayesian estimation, the model error is inferred as well. We have presented numerous numerical examples, showing the capability of handling complex data with incomplete models. We illustrate that nonlinear behavior can be captured quite accurately already with a linear material model. The material model selection within the framework of statFEM is introduced, and the RMSE metric \Rev{and Bayes Factor are} utilized to evaluate the quality of the inferred displacement. An advantage of the statFEM framework is that only three hyperparameters have to be estimated online, while the prior stochastic problem can be solved once in an offline stage. Therefore, it can provide a promising tool for digital twinning, see, e.g., \cite{febrianto2022digital}.

\section*{Declaration of competing interest}
The authors declare that they have no known competing financial interests or personal relationships that could have appeared to influence the work reported in this paper.
\section*{Acknowledgement}\label{sec:Ackn}
The support of the German Research Foundation is gratefully acknowledged in the following projects:
\begin{itemize}
	\item DFG GRK2075-2: \textit{Modelling the constitutional evolution of building materials and structures with respect to aging}.
	\item DFG 501798687: \textit{Monitoring data-driven life cycle management with AR based on adaptive, AI-supported corrosion prediction for reinforced concrete structures under combined impacts}. Subproject of SPP 2388: \textit{Hundred plus - Extending the Lifetime of Complex Engineering Structures through Intelligent Digitalization}.
\end{itemize}
In both projects, methods are developed to link measurement data and physical models to improve structural health monitoring of aging materials and structures.
\section*{Data availability}
Upon acceptance of this manuscript, the accompanying code will be published on zenodo.org.
\numberwithin{equation}{section}
\begin{appendices}
	\section{Probabilistic-type of Hermite Polynomials}\label{Appendix:probabilisticHermitie}
	For the sake of clarification, the univariate Hermite polynomials up to $p=3$ are given as follows:
	\begin{equation}
		\psi_0(\xi_1) = 1, \quad \psi_1(\xi_1) =\xi_1, \quad \psi_2(\xi_1) =\frac{1}{\sqrt[]{2}}(\xi_1^2-1), \quad \psi_3(\xi_1) =\frac{1}{\sqrt[]{6}}(\xi_1^3-3\xi_1)
		\label{eq:univariatePC}
	\end{equation}
	and the multivariate polynomial of second order $p=2$ for two random variables $M=2$ are
	\begin{equation}
		\begin{split}
			&\mPsi_0(\vecxi) = \psi_0(\xi_1) . \psi_0(\xi_2) = 1,          \quad\,\,\,\,\,\,\, 	\mPsi_1(\vecxi) = \psi_1(\xi_1) . \psi_0(\xi_2) = \xi_1\\
			&\mPsi_2(\vecxi) = \psi_0(\xi_1) . \psi_1(\xi_2) = \xi_2,      \quad\,\,\,\,\,	\mPsi_3(\vecxi) = \psi_2(\xi_1) . \psi_0(\xi_2) = \frac{1}{\sqrt[]{2}}(\xi_1^2-1)\\
			&\mPsi_4(\vecxi) = \psi_1(\xi_1) . \psi_1(\xi_2) = \xi_1\xi_2, \quad	\mPsi_5(\vecxi) = \psi_0(\xi_1) . \psi_2(\xi_2) = \frac{1}{\sqrt[]{2}}(\xi_2^2-1).
		\end{split}
		\label{eq:Psi_p2M2}
	\end{equation}

	\section{Matrix Notation of Stochastic Constitutive Tensor}\label{Appendix:MVConstitutive}
	The matrix notation of fourth-order constitutive tensor $\boldsymbol{\mathcal{D}}$ allows us to formulate the constitutive matrix $\vecD$ based on Young's modulus and Poisson's ratio more easily. This notation is extra helpful in this contribution because the uncertainty in the material parameter reflects only in Young's modulus $E$.
	\subsection{Three-dimensional Case}
	For three-dimensional problems, the constitutive matrix $\vecD$ is defined as follows:
	\begin{equation}
		\vecD = E \cdot \vecD^* = E \cdot \frac{1}{(1+\nu)(1-2\nu)}
		\begin{bmatrix}
			1-\nu      & \nu   & \nu   & 0                & 0                & 0                \\
			           & 1-\nu & \nu   & 0                & 0                & 0                \\
			           &       & 1-\nu & 0                & 0                & 0                \\
			           &       &       & \frac{1-2\nu}{2} & 0                & 0                \\
			           &       &       &                  & \frac{1-2\nu}{2} & 0                \\
			\text{sym} &       &       &                  &                  & \frac{1-2\nu}{2}
		\end{bmatrix}.
		\label{eq:3D_C}
	\end{equation}
	The stochastic constitutive matrix based on the stochastic Young's modulus in \eqref{eq:material_PC} is then given by
	\begin{equation}
		\vecD_{P_E}\big(\xi(\theta) \big) = E_{P_E} \big(\xi(\theta) \big) \cdot \vecD^* = \frac{\sum_{i=0}^{P_E} \mu_E \frac{\sigma_\kappa^i}{i !} \psi_i\big(\xi(\theta)\big)}{(1+\nu)(1-2\nu)}
		\begin{bmatrix}
			1-\nu      & \nu   & \nu   & 0                & 0                & 0                \\
			           & 1-\nu & \nu   & 0                & 0                & 0                \\
			           &       & 1-\nu & 0                & 0                & 0                \\
			           &       &       & \frac{1-2\nu}{2} & 0                & 0                \\
			           &       &       &                  & \frac{1-2\nu}{2} & 0                \\
			\text{sym} &       &       &                  &                  & \frac{1-2\nu}{2}
		\end{bmatrix}.
		\label{eq:3D_CStochastic}
	\end{equation}
	\subsection{Two-dimensional Case}
	The stochastic constitutive matrix for plain strain considered in this paper has the following form
	\begin{equation}
		\vecD_{P_E}\big(\xi(\theta) \big) = E_{P_E} \big(\xi(\theta) \big) \cdot \vecD^* =  \frac{\sum_{i=0}^{P_E} \mu_E \frac{\sigma_\kappa^i}{i !} \psi_i\big(\xi(\theta)\big)}{(1+\nu)(1-2\nu)}
		\begin{bmatrix}
			1-\nu      & \nu   & 0                \\
			           & 1-\nu & 0                \\
			\text{sym} &       & \frac{1-2\nu}{2}
		\end{bmatrix},
		\label{eq:2D_CStochastic}
	\end{equation}

	\section{Estimation of Hyperparameters}\label{Appendix:hyperparameterEstimation}
	To estimate the hyperparameters, we use the Bayesian formwork again as follows:
	\begin{equation}
		\prb(\vecw | \yObsOne, \yObsTwo, \dots, \yObsNrep) = \prb(\vecw | \YObs) = \frac{\prb(\YObs | \vecw)\prb(\vecw)}{\int \prb(\YObs |\vecw)\prb(\vecw) \d{\vecw}}.
		\label{eq:statFEM_HyperparametersBayesian}
	\end{equation}
	Here, $\prb(\vecw)$ conceals any prior information about the hyperparameters. In this contribution, we assume a noninformative prior, i.e., $\prb(\vecw) =  1$. The denominator is a constant, which can be neglected if a point estimate is needed. Therefore, \eqref{eq:statFEM_HyperparametersBayesian} can be represent as
	\begin{equation}
		\prb(\vecw | \YObs) \propto \prb(\YObs | \vecw).
		\label{eq:statFEM_HyperparametersBayesianNormalized}
	\end{equation}
	Note that $\prb(\YObs | \vecw)$ is the marginal likelihood in \eqref{eq:statFEM_Bayesian}. Maximizing the likelihood $\prb(\YObs | \vecw)$ means finding the optimal hyperparameters $\vecw^*$ which are most probable to explain $\yObsOne, \yObsTwo, \dots, \yObsNrep$. It means,
	\begin{equation}
		\vecw^* = \argmax_{\vecw}{\Big(\prb(\YObs |  \vecw)\Big)} = \argmax_{\vecw}{\Big(\prod_{i=1}^{\nrep} \prb(\YObsi | \vecw) \Big)}.
		\label{eq:maximumMarginal}
	\end{equation}
	The probability of observation data given hyperparameters has the following form:
	\begin{equation}
		\prb(\YObs |  \vecw) = \Bigg(\frac{1}{\sqrt[]{(2\pi)^{\nsen} \de(\vecvarSigma)}}\Bigg)^{\nrep} \exp{- \onetwo \sum_{i=1}^{\nrep} \Big( (\YObsi-\rho \vecH \MuPC)^T \vecvarSigma^{-1} (\YObsi-\rho \vecH \MuPC)\Big)},
		\label{eq:marginalLikelihoodFull}
	\end{equation}
	with $\vecvarSigma = \CD + \CE + \rho^2 \vecH \CuPC \vecH^T$. Maximizing the natural logarithm of marginal likelihood is preferable to improve numerical stability. Then we have
	\begin{equation}
		\ln{\prb(\YObs |  \vecw)} = -\vecvartheta(\rho,\sigd,\ld)
		\label{eq:marginalLikelihoodLog}
	\end{equation}
	where the function $\vecvartheta$ is given by
	\begin{equation}
		\vecvartheta(\rho,\sigd,\ld) = \onetwo \Bigg(\nrep \nsen\ln{2\pi}+ \nrep \ln{\de{\vecvarSigma}} + \sum_{i=1}^{\nrep} \Big((\YObsi-\rho \vecH \MuPC)^T \vecvarSigma^{-1} (\YObsi-\rho \vecH \MuPC) \Big)\Bigg).
		\label{eq:thetaFun}
	\end{equation}
	Instead of maximizing the negative log-likelihood, we can minimize the log-likelihood. Therefore, \eqref{eq:maximumMarginal} can be reformulated as
	\begin{equation}
		\vecw^* = \argmax_{\rho,\sigd,\ld}{\Big(\ln{\prb(\YObs |  \vecw)}\Big)} = \argmin_{\rho,\sigd,\ld} \vecvartheta(\rho,\sigd,\ld).
		\label{eq:minMarginal}
	\end{equation}
	In order to fulfill the numerical stability, the hyperparameters $\sigd$ and $\ld$ are also transformed to natural logarithmic space $\lsigd$ and $\lld$, respectively. Consequently, the kernel function in \eqref{eq:SE_Kernel_d} is given by
	\begin{equation}
		c_\vecd(\vecX,\vecX') = \expbig{2\lsigd}\exp{-\frac{||\vecX-\vecX'||^2}{2}\expbig{-2\lld}}.
		\label{eq:SE_Kernel_d_log}
	\end{equation}
	Moreover, its partial derivatives w.r.t $\lsigd$ and $\lld$ are given by:
	\begin{equation}
		\dcsigd(\vecX,\vecX') := \parder{c_\vecd}{\lsigd}   = 2c_\vecd
		\label{eq:deriv_SE_Kernel_sigd_log}
	\end{equation}
	and
	\begin{equation}
		\dcld(\vecX,\vecX') := \parder{c_\vecd}{\ld}   = c_\vecd ||\vecX-\vecX'||^2 \expbig{-2\lld}.
		\label{eq:deriv_SE_Kernel_ld_log}
	\end{equation}
	Note that the corresponding matrices of $\dcsigd(\vecX,\vecX')$, $\dcld(\vecX,\vecX')$ are $\dCsigd$ and $\dCld$, respectively. The partial derivation of \eqref{eq:thetaFun} w.r.t $\rho$, $\lsigd$ and $\lld$ are given by
	\begin{equation}
		\begin{split}
			\parder{\vecvartheta}{\rho} = \onetwo \Bigg(&2 \rho \nrep \tr{\vecvarSigma^{-1} \vecH \CuPC \vecH^T}  -  \sum_{i=1}^{\nrep} \bigg(  (\vecH \MuPC)^T \vecvarSigma^{-1} (\YObsi - \rho \vecH \MuPC) + \\
			& 2 \rho (\YObsi - \rho \vecH \MuPC)^T  (\vecvarSigma^{-1} \vecH \CuPC \vecH^T \vecvarSigma^{-1}) (\YObsi - \rho \vecH \MuPC) + \\
			&  (\YObsi - \rho \vecH \MuPC)^T \vecvarSigma^{-1}  (\vecH \MuPC) \bigg) \Bigg),
		\end{split}
		\label{eq:deriv_thetaFun_rho}
	\end{equation}
	\begin{equation}
		\parder{\vecvartheta}{\lsigd} = \onetwo \Bigg(\nrep \tr{\vecvarSigma^{-1} \dCsigd}  -  \sum_{i=1}^{\nrep} \bigg( (\YObsi - \rho \vecH \MuPC)^T  (\vecvarSigma^{-1}  \dCsigd \vecvarSigma^{-1}) (\YObsi - \rho \vecH \MuPC) \bigg) \Bigg)
		\label{eq:deriv_thetaFun_sigd}
	\end{equation}
	and
	\begin{equation}
		\parder{\vecvartheta}{\lld} = \onetwo \Bigg(\nrep \tr{\vecvarSigma^{-1} \dCld}  -  \sum_{i=1}^{\nrep} \bigg( (\YObsi - \rho \vecH \MuPC)^T  (\vecvarSigma^{-1}  \dCld \vecvarSigma^{-1}) (\YObsi - \rho \vecH \MuPC) \bigg) \Bigg)
		\label{eq:deriv_thetaFun_ld}
	\end{equation}
	respectively. In practice,  the two terms $\vecvarSigma^{-1}$ and $\de{\vecvarSigma}$ are computationally expensive to evaluate. Therefore,  the minimization has to be implemented based on the Cholesky decomposition of $\vecvarSigma$. The optimization \eqref{eq:minMarginal} given its partial derivatives is then formulated as follows:
	\begin{equation}
		\begin{split}
			& (\rho^*, \lsigdStr, \lldStr ) = \argmin_{\rho,\lsigd,\lld} \vecvartheta(\rho,\sigd,\ld) \\ \\
			& \text{such that:} \,\, \parder{\vecvartheta}{\rho} = 0, \,\, \parder{\vecvartheta}{\lsigd} = 0, \,\, \parder{\vecvartheta}{\lld} =0.
		\end{split}
		\label{eq:minMarginalDerivative}
	\end{equation}
\end{appendices}

\bibliographystyle{unsrtnat}
\bibliography{literature}
\end{document}